\documentclass{amsart}

\usepackage{amscd,amsmath,xypic,amssymb,combelow,tikz-cd,etoolbox,calligra,mathrsfs,enumitem,mathtools,hyperref,float,bm}

\makeatletter
\patchcmd{\@settitle}{\uppercasenonmath\@title}{}{}{}
\makeatother

\xyoption{all}
\CompileMatrices

\makeatletter
\@addtoreset{equation}{section}
\makeatother

\newtheorem{theorem}[subsection]{Theorem}

\newtheorem{proposition}[subsection]{Proposition}
\newtheorem{lemma}[subsection]{Lemma}
\newtheorem{corollary}[subsection]{Corollary}

\newtheorem{definition}[subsection]{Definition}
\newtheorem{assumption}[subsection]{Assumption}

\newtheorem{claim}[subsection]{Claim}

\newtheorem{remark}[subsection]{Remark}

\def\loccit{\emph{loc. cit. }}

\def\fgl{{\mathfrak{gl}}}

\def\BC{{\mathbb{C}}}
\def\BK{{\mathbb{K}}}
\def\BL{{\mathbb{L}}}

\def\BN{{\mathbb{N}}}

\def\BR{{\mathbb{R}}}
\def\BQ{{\mathbb{Q}}}
\def\BT{{\mathbb{T}}}
\def\BZ{{\mathbb{Z}}}

\def\CS{{\mathcal{S}}}

\def\CV{{\mathcal{V}}}

\def\ev{\textrm{ev}}

\def\ev{\textrm{ev}}

\def\vs{\varsigma}

\def\and{\textrm{ }\&\textrm{ }}

\def\sym{\textrm{sym}}
\def\Sym{\textrm{Sym}}

\def\esym{\emph{sym}}
\def\eSym{\emph{Sym}}

\def\oCS{\mathring{\CS}}
\def\tCS{\boldsymbol{\dot}{\CS}}

\def\nn{{{\BN}}^I}

\def\UU{\mathbf{U}}
\def\UUo{\mathbf{U}^0}
\def\UUp{\mathbf{U}^+}
\def\UUpm{\mathbf{U}^\pm}
\def\UUm{\mathbf{U}^-}

\def\tUU{{\widetilde{\mathbf{U}}}}
\def\tUUp{{\widetilde{\mathbf{U}}^+}}
\def\tUUpm{{\widetilde{\mathbf{U}}^\pm}}

\def\tUUm{{\widetilde{\mathbf{U}}^-}}

\def\bs{{\boldsymbol{\vs}}}

\def\b0{{\boldsymbol{0}}}

\def\bn{\boldsymbol{n}}

\def\op{\text{op}}

\def\tUpsilon{\widetilde{\Upsilon}}
\def\tzeta{\widetilde{\zeta}}

\def\tQ{\widetilde{Q}}

\begin{document}

\title[Reduced quiver quantum toroidal algebras]{\Large{\textbf{Reduced quiver quantum toroidal algebras}}}

\author[Andrei Negu\cb t]{Andrei Negu\cb t}
\address{MIT, Department of Mathematics, Cambridge, MA, USA \newline 
	\text{ } \ \ Simion Stoilow Institute of Mathematics, Bucharest, Romania}
\email{andrei.negut@gmail.com}

\maketitle

\begin{abstract} We give a generators-and-relations description of the reduced versions of quiver quantum toroidal algebras, which act on the spaces of BPS states associated to (non-compact) toric Calabi-Yau threefolds $X$. As an application, we obtain a description of the $K$-theoretic Hall algebra of (the quiver with potential associated to) $X$, modulo torsion.

\end{abstract}

\section{Introduction}
\label{sec:intro}

\medskip

\subsection{} 
\label{sub:physics}

Let $X$ be a (non-compact) toric Calabi-Yau threefold. To $X$ one can associate a 2d quantum field theory with four supercharges, and we will be interested in two features of this theory: its vector space of BPS states, and more importantly for us, the BPS algebra which acts on said vector space. The latter algebra has been dubbed the quiver quantum toroidal algebra (\cite{GLY1,GLY2,NW1,NW2}, following \cite{LY}).

\medskip

\noindent Before we dive into the definition of the quiver quantum toroidal algebra $\tUU$, let us recall certain objects associated to the Calabi-Yau threefold $X$
$$
X \leadsto \text{toric diagram} \leadsto \text{brane tiling} \leadsto \text{quiver}
$$
We refer the reader to \cite[Appendix C]{NW2} for a detailed review of the procedures $\leadsto$ listed above, and we simply state the following properties of the objects involved. 

\medskip

\begin{figure}[h]
\includegraphics[scale=0.45]{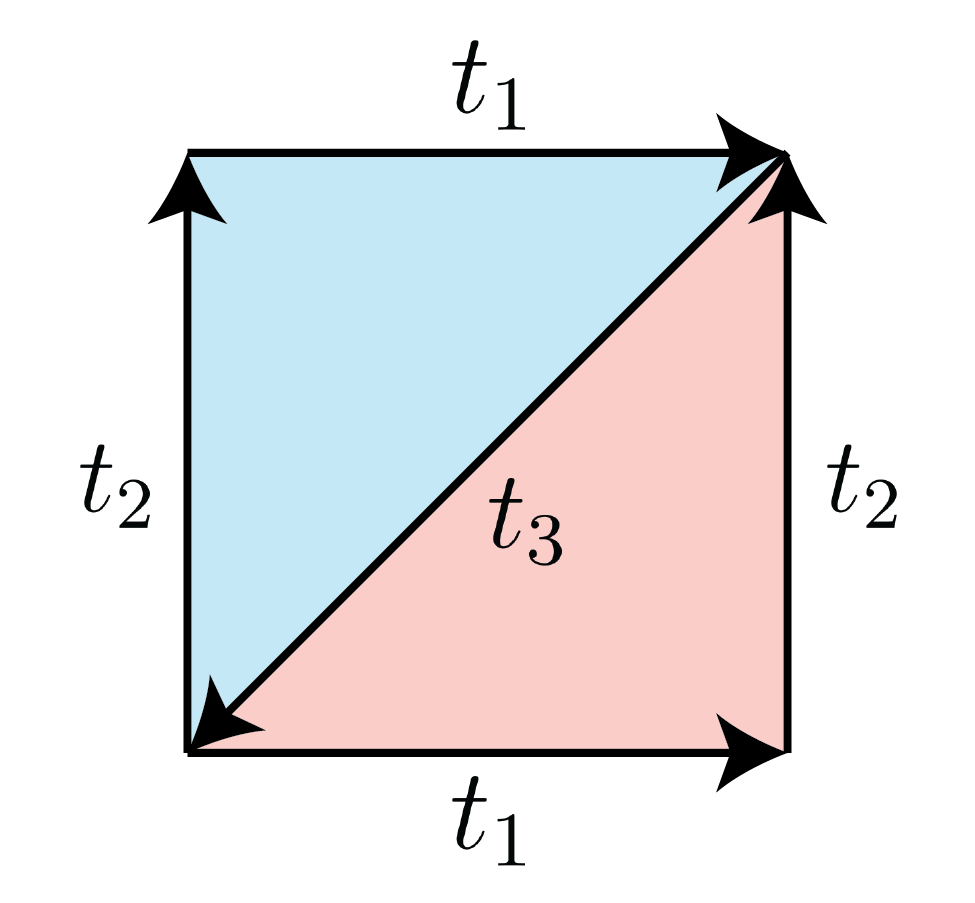} 
\caption{The quiver associated to $X = \BC^3$. The square is the usual representation of the flat torus, so the quiver has one vertex, three edges and two faces.}
\label{fig:1}
\end{figure}

\begin{itemize}[leftmargin=*]

\item The toric diagram associated to $X$ is a particular collection of points in $\BZ^2$ and line segments between them. 

\medskip

\item The normals to the aforementioned line segments can be drawn on the torus $\BT^2$, and they define a brane tiling, i.e. a decomposition of the torus into polygonal regions called faces. Very importantly, the faces can be colored in blue and red such that any two faces which share an edge have different colors. \footnote{As just described, the brane tiling yields a graph $G$ drawn on the torus. In the literature, the term ``brane tiling" is sometimes applied to the dual graph of $G$, which is bipartite.}

\medskip

\item The vertices and edges of the aforementioned faces determine a quiver $Q$ drawn on $\BT^2$. The bicolorability property of the brane tiling implies that the edges of $Q$ can be oriented so that they go clockwise around the blue faces and counterclockwise around the red faces. The interested reader may find the quiver associated to the Calabi-Yau threefold $X = \BC^3$ in Figure \ref{fig:1}.

\end{itemize}

\medskip

\subsection{}
\label{sub:consistent}

As the definition of the quiver quantum toroidal algebra $\tUUp$ only takes the quiver as input, one can state the construction in generality greater than those quivers which arise from toric Calabi-Yau threefolds via the procedure above.

\medskip

\begin{definition}
\label{def:quiver intro}

Let $Q$ be a quiver drawn on a torus (with vertex set $I$ and edge set $E$), whose faces are colored in blue and red such that the two incident faces to a given edge have different colors. We assume that the edges of the quiver are oriented so as to go clockwise around the blue faces.

\end{definition}

\medskip

\noindent We will let $ \pi : \tQ \rightarrow Q$ denote the lift of the aforementioned quiver to the universal cover $\BR^2 \rightarrow \BT^2$, and note that $\tQ$ inherits the blue and red colored faces of $Q$. In the present paper, ``paths" and ``cycles" in a quiver will refer to the oriented notions.

\medskip

\begin{definition}
\label{def:quasi}

A \textbf{broken wheel} refers to a path obtained by removing a single edge $e$ from the boundary of any face $F$ of $\tQ$. The \textbf{mirror image} of the aforementioned broken wheel is the path obtained by removing $e$ from the boundary of the other face $F' \neq F$ incident to $e$. The edge $e$ will be called the \textbf{interface} of the broken wheel (and of its mirror image).

\begin{figure}[h]
\includegraphics[scale=0.45]{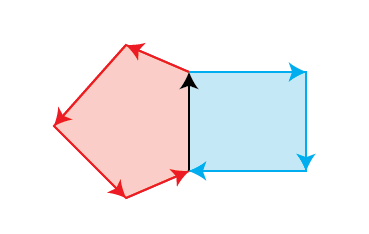} 
\caption{A broken wheel (the path in red) and its mirror image (the path in blue). The black arrow is the interface.}
\label{fig:2}
\end{figure}

\end{definition}

\medskip

\begin{definition}
\label{def:consistent intro}

The quiver $Q$ is called \textbf{shrubby} if given any paths $p \neq p'$ in $\tQ$ with the same start and end points, at least one of $p$ and $p'$ contains a broken wheel whose interface lies in the closed region between the two paths. 

\end{definition}

\medskip

\noindent When one of $p$ and $p'$ is trivial, Definition \ref{def:consistent intro} states that any cycle in $\tQ$ must contain a broken wheel in the closure of its interior. We will see in Lemma \ref{lem:non-degenerate is shrubby} that the shrubbiness condition above is implied by more traditional notions of consistency of brane tilings and dimer models, such as the existence of a non-degenerate $R$-charge. We do not know (and it is an interesting question) whether all quivers which arise from Calabi-Yau threefolds as in Subsection \ref{sub:physics} are shrubby.

\medskip

\subsection{}
\label{sub:zeta}

Let $\BK$ be a field of characteristic 0. To every edge $e$ of the quiver $Q$, we associate a \textbf{parameter} $t_e \in \BK^\times$ such that for every face $F$ of $Q$ we have \footnote{In the setting of toric Calabi-Yau threefolds $X$, one usually takes $\BK = \BQ(q_1,q_2)$ where $q_1,q_2$ are elementary characters of the rank 2 torus that acts on $X$ by preserving the Calabi-Yau 3-form. In this setting, the parameters $t_e$ are monomials in $q_1$ and $q_2$.}
\begin{equation}
\label{eqn:loop constraint}
t_{e_1}\dots t_{e_k} = 1
\end{equation}
where $e_1,\dots,e_k$ is the boundary of $F$. Fix positive real numbers
\begin{equation}
\label{eqn:abs}
\{ |t_e| \}_{e \text{ edge}} \subset \BR_{>0}
\end{equation}
such that $|t_{e_1}|\dots |t_{e_k}|=1$ for any face $F$, with the notation as in \eqref{eqn:loop constraint}.

\medskip

\begin{assumption}
\label{assumption}

For any pair of paths 
\begin{equation}
\label{eqn:pair of paths}
i \xleftarrow{e_1} \dots \xleftarrow{e_k} j \xrightarrow{e'_\ell} \dots \xrightarrow{e'_1} i'
\end{equation}
in $\tQ$ such that $i \neq i'$ but $\pi (i) = \pi(i') \in Q$, we have
\begin{equation}
\label{eqn:generic 1}
|t_{e_1}| \dots |t_{e_k}| \neq |t_{e'_1}| \dots |t_{e'_{\ell}}| 
\end{equation} 

\end{assumption}


\medskip

\noindent We expect all the results in the present paper to hold under the weaker assumption that $t_{e_1}\dots t_{e_k} \neq t_{e'_1} \dots t_{e'_{\ell}}$ instead of \eqref{eqn:generic 1}, but we prefer the apparently stronger condition in Assumption \ref{assumption} due to the way our proof of Proposition~\ref{prop:main} is formulated. In \cite{LY} and related works, products of the parameters $t_e$ along paths are interpreted as coordinate functions of atoms in BPS crystals; property \eqref{eqn:generic 1} is closely related to the requirement that different atoms have different coordinates.

\medskip

\noindent The edge parameters can be assembled into the following rational functions
\begin{equation}
\label{eqn:def zeta}
\zeta_{ij}(x) = \frac {\alpha_{ij} x^{s_{ij}}}{(1-x)^{\delta_{ij}}} \prod_{e : i \rightarrow j} (1-xt_e) \in \BK(x)
\end{equation}
for all $i,j \in I$, where $\alpha_{ij} \in \BK^\times$ and $s_{ij} \in \BZ$ are suitably chosen (but will not play an important role in the present paper, so we will not specify them explicitly). 

\medskip

\begin{remark}
\label{rem:intro}

Moreover, different authors use different conventions on $\alpha_{ij}$ and $s_{ij}$. For example, \cite{GLY1} requires $s_{ij}$ to be minus half the number of arrows from $i$ to $j$; this situation can also be accommodated by the present paper, at the cost of replacing polynomials built out of integer powers by polynomials built out of half-integer powers. We will avoid this setup in order to not overburden our notation.

\end{remark}

\medskip

\subsection{}
\label{sub:qqta}

Using the data in Subsection \ref{sub:zeta}, we will now review the definition of the quantum toroidal algebra associated to the quiver $Q$ and parameters $\{t_e\}_{e \in E}$, which was introduced in \cite{GLY1,NW1} as a trigonometric version of the quiver Yangian of \cite{LY} (see also \cite{RSYZ} for a closely related mathematical construction). 

\medskip

\begin{definition}
\label{def:quad intro}

The (half) \textbf{quiver quantum toroidal algebra} $\tUUp$ is
\begin{equation}
\label{eqn:def positive}
\tUUp = \BK \Big\langle e_{i,d} \Big \rangle_{i \in I, d \in \BZ} \Big / \text{relation \eqref{eqn:rel quad}}
\end{equation}
where if we write
$$
e_i(z) = \sum_{d \in \BZ} \frac {e_{i,d}}{z^d}
$$
then the defining relations are given by the formula
\begin{equation}
\label{eqn:rel quad}
e_i(z) e_j(w) \zeta_{ji} \left(\frac wz\right) = e_j(w) e_i(z) \zeta_{ij} \left( \frac zw \right)
\end{equation}
for all $i,j \in I$. \footnote{Relation \eqref{eqn:rel quad} is interpreted as an infinite collection of relations obtained by equating the coefficients of all $\{z^aw^b\}_{a,b\in \BZ}$ in the left and right-hand sides (if $i = j$, one clears the denominators $z-w$ from \eqref{eqn:rel quad} before equating coefficients).}

\end{definition}

\medskip

\noindent Define $\tUUm = \tUU^{+,\text{op}}$, and denote its generators by $f_{i,d}$ instead of $e_{i,d}$. Finally, let us consider the commutative algebra
$$
\UUo = \BK\left[h_{i,d}, h'_{i,d'}\right]_{i \in I, d,d' \geq \text{appropriately chosen integers}} 
$$
Then the (full) \textbf{quiver quantum toroidal algebra} is defined as
\begin{equation}
\label{eqn:full}
\tUU = \tUUp \otimes \UUo \otimes \tUUm
\end{equation}
with certain commutation relations imposed between elements in the three tensor factors above. We refer the reader to \cite{GLY1, NW1} for the explicit commutation relations, as they will not be used in the present paper; instead, we will only focus on $\tUUp$. 


\medskip

\subsection{}
\label{sub:action}

The main motivation for defining the algebra $\tUU$ is that it acts on the vector space of so-called BPS crystal configurations
\begin{equation}
\label{eqn:action intro}
\tUU \curvearrowright M = \bigoplus_{\Lambda \text{ 3d crystal configuration}} \BK \cdot |\Lambda\rangle  
\end{equation}
(see \cite[Section 5]{NW1} for a review of 3d crystal configurations, which are generalizations of plane partitions). We will not make the action \eqref{eqn:action intro} explicit, so we will not make any rigorous claims about it and merely use it as motivation for our subsequent constructions. The main goal of the present paper is to describe the kernel of the action \eqref{eqn:action intro}, i.e. to define the smallest possible quotient
\begin{equation}
\label{eqn:quotient basic}
\tUU \twoheadrightarrow \UU
\end{equation}
such that the action \eqref{eqn:action intro} factors through an action of $\UU$. To this end, we will consider the \textbf{shuffle algebra} realization of quiver quantum toroidal algebras
$$
\tUpsilon^\pm : \tUUpm \longrightarrow \CV^\pm = \bigoplus_{\bn \in \BN^I} \BK[z_{i1},z_{i1}^{-1},\dots,z_{in_i}, z_{in_i}^{-1}]_{i \in I}^{\sym}
$$
(we refer the reader to Subsection \ref{sub:def shuf} for a description of the shuffle product on $\CV^\pm$, and to Subsection \ref{sub:upsilon} for the definition of the homomorphism $\tUpsilon^\pm$). Set
\begin{equation}
\label{eqn:quotient intro}
\UUpm = \tUUpm \Big / \text{Ker }\tUpsilon^{\pm}
\end{equation}
As noted in \cite[Section 5]{GLY1}, the action \eqref{eqn:action intro} factors through the shuffle algebra. Therefore, the \textbf{reduced} (full) quiver quantum toroidal algebra
\begin{equation}
\label{eqn:reduced intro}
\UU = \UUp \otimes \UUo \otimes \UUm
\end{equation}
will inherit an action on $M$ from \eqref{eqn:action intro}. To define this action, one needs to impose the same commutation relations between the tensor factors of \eqref{eqn:reduced intro} as between the tensor factors of \eqref{eqn:full}. We will not present these relations explicitly in the present paper, and make no rigorous claims about them. Instead, we will focus on $\UUp$. 

\medskip

\subsection{}

\noindent The main purpose of the present paper is to describe $\UU^\pm$ by explicitly presenting the quotient \eqref{eqn:quotient intro}. More specifically, we will describe a collection of generators for the two-sided ideal $\text{Ker } \tUpsilon^\pm$. For every face $F = \{i_0,i_1,\dots,i_{k-1},i_k=i_0\}$ of the quiver $Q$ (note that some of the indices $i_0,\dots,i_{k-1}$ may be repeated within a given face), consider the following parameters corresponding to the edges of $F$
\begin{equation}
\label{eqn:cycle intro}
t_a = t_{\overrightarrow{i_{a-1}i_a}}
\end{equation}
Note that $t_1\dots t_k = 1$ due to \eqref{eqn:loop constraint}. Let $\tzeta_{ij}(x) = \zeta_{ij}(x) (1-x)^{\delta_{ij}}$
for all $i,j \in I$. We define the formal series
\begin{equation}
\label{eqn:series intro}
e_F(x_1,\dots,x_k) \in \tUUp[[x_1, x_1^{-1}, \dots, x_k,x_k^{-1}]]
\end{equation}
by the following formula
\begin{multline}
\sum_{a=1}^k \frac {x_1t_2\dots t_a}{x_a} \cdot \frac { \prod_{b \succ c} \tzeta_{i_ci_b} \left(\frac {x_c}{x_b} \right)\left( - \frac {x_b}{x_c} \right)^{\delta_{i_bi_c} \delta_{b<c}} }{\prod_{b \sim c + 1} \left(1 - \frac {x_ct_b}{x_b} \right)} \cdot \\ \cdot e_{i_{a}}(x_{a}) \dots e_{i_1}(x_1) e_{i_k}(x_k)  \dots  e_{i_{a+1}}(x_{a+1}) \label{eqn:formula series intro}
\end{multline}
In \eqref{eqn:formula series intro}, the notation $b \succ c$ (respectively $b \sim c+1$) means that $b$ precedes (respectively immediately precedes) $c$ in the sequence $(a,\dots,1,k,\dots,a+1)$. The symbols $\delta_{b<c}$ and $\delta_{i_bi_c}$ are defined as in Subsection \ref{sub:toric cy}. Note that the first line of \eqref{eqn:formula series intro} is a Laurent polynomial in $x_1,\dots,x_k$, due to the fact that all the denominators 
$$
1-\frac {x_ct_b}{x_b}
$$
are canceled by the $\tzeta$ functions in the numerator. The following is our main result. 

\bigskip

\begin{theorem}
\label{thm:main}

If $Q$ is shrubby (as in Definition \ref{def:consistent intro}), then the coefficients of the series \eqref{eqn:series intro} generate $\emph{Ker } \tUpsilon^+$ as a two-sided ideal. In other words, we have
\begin{equation}
\label{eqn:relations intro}
\UUp = \tUUp \Big / \Big(\text{series coefficients of }e_F(x_1,\dots,x_k) \Big)_{F \text{ face of }Q}
\end{equation}
Similar results hold for $\UUm$, by replacing $e$'s with $f$'s and reversing the order of the factors in the product on the second line of \eqref{eqn:formula series intro}. \footnote{While the quotient \eqref{eqn:relations intro} imposes a $\BZ^k$-worth of relations for every face $F$ with $k$ vertices, we will see in Remark \ref{rem:finite} that these can be reduced to a $\BZ$-worth of relations for every face. More precisely, arbitrarily choosing one non-zero coefficient of the series $e_F$ in each integer homogeneous degree instead of all coefficients (for every face $F$) would determine the same quotient in \eqref{eqn:relations intro}.}

\end{theorem}

\bigskip

\noindent Lemma \ref{lem:non-degenerate is shrubby} implies that a large family of physically interesting Calabi-Yau threefolds $X$ correspond to shrubby quivers, and so Theorem \ref{thm:main} applies to them. We conclude that the relations which we factor in \eqref{eqn:relations intro} are the sought-for ``Serre relations" of \cite{GLY1}. The terminology of these relations is historically motivated by the analogous situation of quantum loop groups associated to finite type Dynkin diagrams, in which the role of relations \eqref{eqn:relations intro} is played by the Drinfeld-Serre relations. Note, however, that the classic Drinfeld-Serre relations are not enough to characterize quantum loop groups associated to general Dynkin diagrams (see \cite{Loop}).

\medskip

\begin{remark}
\label{rem:intro 1}

If $Q$ is not shrubby, then we expect that one needs additional relations besides \eqref{eqn:formula series intro}. In this situation, the ideal $\emph{Ker }\tUpsilon^+$ can be studied according to the general principles of \cite{Arbitrary}, but we do not know explicit generators of this ideal.

\end{remark}

\medskip

\begin{remark}
\label{rem:intro 2}

It is straightforward to write down rational/elliptic versions of the relations \eqref{eqn:formula series intro}, which would give necessary relations that hold in the rational/elliptic counterparts of the reduced algebra $\UUp$ (see \cite{GLY1} for an overview). However, in the rational/elliptic settings, we do not know whether these relations are also sufficient, i.e. if they generate the analogue of the two-sided ideal $\emph{Ker }\tUpsilon^+$.

\end{remark}

\medskip

\subsection{}
\label{sub:examples}

Let us spell out the constructions above in the case $X = \BC^3$, when the quiver is the one in Figure \ref{fig:1}. There is a single vertex, so $I = \{\bullet\}$ and we will henceforth suppress the index $\bullet \in I$ from all our formulas. There are three edges, whose associated parameters $t_1,t_2,t_3$ satisfy the equation
$$
t_3 = \frac 1{t_1t_2}
$$
We take the ground field to be $\BK = \BQ(t_1,t_2)$. The only $\zeta$ function \eqref{eqn:def zeta} is
$$
\zeta(x) = \frac {x^{-1}(1-xt_1)(1-xt_2)(1-xt_3)}{1-x}
$$
(the particular choice of the monomial $x^{-1}$ was made in order to match existing conventions in the literature). The (half) quiver quantum toroidal algebra \eqref{eqn:def positive} is generated by a single formal series $e(z)$ modulo the quadratic relation
\begin{multline}
\label{eqn:c3 quad}
e(z) e(w)  (z-wt_1)(z-wt_2)(z-wt_3) = \\ =  e(w) e(z) (zt_1-w)(zt_2-w)(zt_3-w)
\end{multline}
Meanwhile, formula \eqref{eqn:formula series intro} for the red face in Figure \ref{fig:1} reads
\begin{equation}
\label{eqn:c3 cubic}
\begin{split} 
e_{F_{\text{red}}}(x_1,x_2,x_3) = \frac {\prod_{i=1}^3 [ (x_1-x_2t_i)(x_1-x_3t_i)(x_3-x_2t_i) ]}{x_1x_2^3x_3^3(x_1-x_3t_1)(x_3-x_2t_3)} e(x_1)e(x_3)e(x_2)  \\
+ \frac {t_2\prod_{i=1}^3 [ (x_2-x_1t_i)(x_1-x_3t_i)(x_2-x_3t_i) ]}{x_2^3x_3^4(x_2-x_1t_2)(x_1-x_3t_1)} e(x_2)e(x_1)e(x_3)  \\
+ \frac {t_2t_3\prod_{i=1}^3 [ (x_2-x_1t_i)(x_3-x_1t_i)(x_3-x_2t_i) ]}{x_1x_2^2x_3^4(x_3-x_2t_3)(x_2-x_1t_2)} e(x_3)e(x_2)e(x_1)
\end{split}
\end{equation}
while the analogous expression $e_{F_{\text{blue}}}$ for the blue face is obtained by replacing $\{t_1,t_2,t_3\} \leftrightarrow \{t_3,t_2,t_1\}$. Note that the expressions in $x_1,x_2,x_3$ that precede the series $e(\dots)$ in the three lines of \eqref{eqn:c3 cubic} are actually Laurent polynomials, and so it makes sense to talk about their coefficients. Theorem \ref{thm:main} states that
$$
\UUp = \frac {\BQ(t_1,t_2) \langle \dots, e_{-1},e_0,e_1,\dots \rangle}{\text{equation \eqref{eqn:c3 quad} and coefficients of } e_{F_{\text{red}}}(x_1,x_2,x_3) \text{ and } e_{F_{\text{blue}}}(x_1,x_2,x_3)}
$$
Let us make two observations about formulas \eqref{eqn:c3 cubic}, which apply equally well in the more general context of Theorem \ref{thm:main}. Firstly, as explained in Remark \ref{rem:finite}, many of the coefficients of $e_{F_{\text{red}}}$ and $e_{F_{\text{blue}}}$ are superfluous; we would obtain the same reduced quiver quantum toroidal algebra $\UUp$ if we only imposed relations given by a single coefficient of \eqref{eqn:c3 cubic} of every total homogeneous degree in $x_1,x_2,x_3$. This is because any two such coefficients of the same total homogeneous degree are equivalent to each other up to multiples of the quadratic relation \eqref{eqn:c3 quad}. \\

\noindent Secondly (and perhaps most importantly) there is nothing ``canonical" about the relations in $\UUp$ given by setting the coefficients of $e_{F_{\text{red}}}$ and $e_{F_{\text{blue}}}$ equal to 0, since we would obtain the exact same algebra by adding various multiples of relation \eqref{eqn:c3 quad} to the aforementioned coefficients. For example, if we consider the positive half of the well-known quantum toroidal algebra (\cite{M})
$$
U_{t_1,t_2}^+(\widehat{\widehat{\fgl}}_1) = \frac {\BQ(t_1,t_2) \langle \dots, e_{-1},e_0,e_1,\dots \rangle}{\text{equation \eqref{eqn:c3 quad} and } [[e_{k+1},e_{k-1}],e_k]=0, \ \forall k \in \BZ}
$$
then we have an isomorphism 
$$
\UUp \cong U_{t_1,t_2}^+(\widehat{\widehat{\fgl}}_1), \quad e_k \mapsto e_k, \ \forall k \in \BZ
$$
on account of the fact that both algebras are isomorphic to the shuffle algebra $\CS^+$ of Section \ref{sec:shuffle} (see Theorem \ref{thm:equal} and \cite[Theorem 7.3]{Ts}). However, the cubic relations in the two algebras look quite different, and the fact that they can be obtained from each other by adding multiples of \eqref{eqn:c3 quad} is a very involved computation.

\medskip

\subsection{}
\label{sub:coha}

Quiver quantum toroidal algebras are related to the $K$-theoretic Hall algebras (defined in \cite{Pad}, by analogy with the cohomological Hall algebras of \cite{KS})
$$
K(Q,W)
$$
defined with respect to the following potential
$$
W = \sum_{F \text{ face of }Q} (-1)^F \prod_{e \text{ edge around }F} \phi_e \in \BC[Q] 
$$
where $(-1)^F$ is $+1$ or $-1$ depending on whether the face $F$ is blue or red, and the symbols $\phi_e$ denote generators of the path algebra $\BC[Q]$. We consider $K(Q,W)$ as an algebra over the ring $\BL$ of polynomials in the edge parameters (modulo \eqref{eqn:loop constraint}), and let our ground field be $\BK = \text{Frac}(\BL)$. Then the localized $K$-theoretic Hall algebra 
$$
K(Q,W)_{\text{loc}} = K(Q,W) \bigotimes_{\BL} \BK
$$
is endowed with an algebra homomorphism
$$
K(Q,W)_{\text{loc}} \xrightarrow{\iota} \CV^+
$$
By combining Theorem \ref{thm:equal}, Definition \ref{def:shuffle wheel} and Proposition \ref{prop:main}, the image of $\tUpsilon^+$ can be described as the subspace $\CS^+ \subset \CV^+$ of Laurent polynomials \footnote{For any face $F = \{i_1,\dots,i_{k-1},i_k\}$, we use the notation $z_1,\dots,z_k$ to represent variables of $R$ in accordance with \eqref{eqn:relabeling}, i.e. one should interpret $z_a = z_{i_a\bullet_a}$ for certain $\bullet_a \in \BN$, $\forall a \in \{1,\dots,k\}$.}  $R(z_1,\dots,z_k,\dots)$ which vanish whenever their variables are specialized according to the rule
\begin{equation}
\label{eqn:specialization intro}
\Big\{ z_a = z_{a-1} t_a \Big\}_{a \in \{1,\dots,k\}}
\end{equation}
for any face $F$ of $Q$ (in the notation of \eqref{eqn:cycle intro}). This yields the following result.

\medskip

\begin{corollary}
\label{cor:main}

If $Q$ is shrubby, the images of $\iota$ and $\tUpsilon^+$ coincide, i.e. the localized $K$-theoretic Hall algebra surjects onto the subspace $\CS^+ \subset \CV^+$ of Laurent polynomials which vanish when their variables are specialized to \eqref{eqn:specialization intro}, for any face $F$.

\end{corollary} 

\medskip

\begin{proof} The fact that the image of $\tUpsilon^+$ is (tautologically) generated by $\{z_{i1}^d\}_{i \in I, d\in \BZ}$, which all lie in the image of $\iota$, implies that
\begin{equation}
\label{eqn:inclusion intro 1}
\text{Im } \tUpsilon^+ \subseteq \text{Im }\iota
\end{equation}
To prove the opposite inclusion, one needs to show that the image of $\iota$ is contained in the subspace of Laurent polynomials which vanish when their variables are specialized according to \eqref{eqn:specialization intro} for every face $F$. Following an idea of Yu Zhao (\cite{Yu}), we note that the specialization in question can be realized as restriction to the locally closed subset $U$ of quiver representations $(\phi_{e} : \BC^{n_i} \rightarrow \BC^{n_j})_{e : i \rightarrow j}$ whose only non-zero elements are
$$
\phi_{\overrightarrow{i_1i_2}} \in \BC^* E_{\bullet_1\bullet_2}, \dots, \phi_{\overrightarrow{i_{k-1}i_k}} \in \BC^* E_{\bullet_{k-1}\bullet_k}, \phi_{\overrightarrow{i_ki_1}} \in \BC^* E_{\bullet_k\bullet_1}
$$
(where $E_{ab}$ denote the matrix units with respect to the standard basis of $\{\BC^{n_i}\}_{i \in I}$, and the natural numbers $\bullet_1,\dots,\bullet_k$ are chosen as in \eqref{eqn:relabeling}). Since the locally closed subset $U$ does not intersect the zero locus of $\text{Tr}(W)$, on which $K(Q,W)$ is supported, any $\gamma \in \text{Im }\iota$ has the propery that it restricts to zero in the $K$-theory of $U$. However, it is elementary to see that the latter condition is equivalent to the vanishing of $\gamma$ at the specialization \eqref{eqn:specialization intro}. Because of this, we conclude that $\gamma \in \CS^+$ and so we infer the opposite inclusion to \eqref{eqn:inclusion intro 1}:
\begin{equation}
\label{eqn:inclusion intro 2}
\CS^+ = \text{Im } \tUpsilon^+ \supseteq \text{Im }\iota
\end{equation} \end{proof}

\subsection{} The structure of the present paper is the following. 

\medskip

\begin{itemize}[leftmargin=*]

\item In Section \ref{sec:shuffle}, we discuss $\tUUp$ and its shuffle algebra interpretation for general quivers $Q$.

\medskip

\item In Section \ref{sec:consistent}, we study $\tUUp$ for the particular quivers $Q$ of Definition \ref{def:quiver intro}, and prove Theorem \ref{thm:main}.

\medskip

\item In Section \ref{sec:gardening}, we provide some key results on shrubs (which are certain subgraphs of the universal cover of $Q$ that we use in the proof of Theorem \ref{thm:main}).

\end{itemize}

\medskip

\subsection{} I would like to thank Ben Davison, Richard Kenyon and Masahito Yamazaki for very useful conversations about the topics in the present paper. I gratefully acknowledge NSF grant DMS-$1845034$, as well as support from the MIT Research Support Committee.

\bigskip

\section{Shuffle algebras in general} 
\label{sec:shuffle}

\medskip

We will now recall the basic theory of trigonometric shuffle algebras, in the generality of \cite{Arbitrary}. Thus, throughout the present Section, $Q$ will denote an arbitrary quiver (whose vertex and edge sets will be denoted by $I$ and $E$, respectively), $\BK$ will denote an arbitrary field of characteristic zero, and $\zeta_{ij}(x)(1-x)^{\delta_{ij}}$ will denote arbitrary Laurent polynomials with coefficients in $\BK$ for all $i,j \in I$. Throughout the present paper, the set $\BN$ will be thought to contain 0.

\medskip

\subsection{}
\label{sub:def shuf}

Let us consider an infinite collection of variables $z_{i1},z_{i2},\dots$ for all $i \in I$. For any $\bn = (n_i)_{i \in I} \in \nn$, we will write $\bn! = \prod_{i \in I} n_i!$. The following construction is a straightforward generalization of the trigonometric quantum loop groups of \cite{E, FO}.

\medskip

\begin{definition}
\label{def:big shuf}

The \textbf{big shuffle algebra} associated to the datum $\{\zeta_{ij}(x)\}_{i,j \in I}$ is
$$
\CV^+ = \bigoplus_{\bn \in \BN^I} \BK[z_{i1}, z_{i1}^{-1},\dots,z_{in_i}, z_{in_i}^{- 1}]_{i \in I}^{\esym}
$$
endowed with the multiplication
\begin{equation}
\label{eqn:shuf prod}
R(z_{i1},\dots,z_{in_i}) * R'(z_{i1},\dots,z_{in_i'} ) = 
\end{equation}
$$
\eSym \left[ \frac {R(z_{i1},\dots,z_{in_i}) R'(z_{i,n_i+1},\dots,z_{i,n_i+n_i'})}{\bn! \bn'!} \mathop{\prod^{i,j \in I}_{1\leq a\leq n_i}}_{n_j < b \leq n_j+n_j'} \zeta_{ij} \left(\frac {z_{ia}}{z_{jb}} \right) \right]
$$
Above and henceforth, ``\emph{sym}" (resp. ``\emph{Sym}") denotes symmetric functions (resp. symmetrization) with respect to the variables $z_{i1},z_{i2},\dots$ for each $i \in I$ separately. \footnote{Although the $\zeta$ functions might seem to contribute simple poles at $z_{ia}-z_{ib}$ for $a \neq b$ to the right-hand side of \eqref{eqn:shuf prod}, these poles disappear when taking the symmetrization (the poles in question can only have even order in any symmetric rational function).}

\end{definition} 

\medskip

\noindent By defining the subspace $\CV_{\bn} \subset \CV^+$ to consist of rational functions in $\bn = (n_i)_{i \in I}$ variables, we obtain a decomposition
\begin{equation}
\label{eqn:graded pieces}
\CV^+ = \bigoplus_{\bn \in \nn} \CV_{\bn}
\end{equation}
For example, the Laurent polynomial in a single variable $z_{i1}^d$ lies in $\CV_{\bs^i}$, where
$$
\bs^i = (\underbrace{0,\dots,0,1,0,\dots,0}_{1\text{ on }i\text{-th position}}) \in \nn
$$
We will also consider the opposite big shuffle algebra $\CV^- = \CV^{+,\op}$, whose graded components analogous to \eqref{eqn:graded pieces} will be denoted by $\CV_{-\bn}$, for all $\bn \in \nn$. 

\medskip

\subsection{}
\label{sub:upsilon}

Recall that $\tUUp$ is the quiver quantum toroidal algebra of Definition \ref{def:quad intro}, and $\tUUm$ denotes its opposite. There exist $\BK$-algebra homomorphisms
\begin{equation}
\label{eqn:tilde upsilon}
\tUpsilon^\pm : \tUUpm \longrightarrow \CV^{\pm}, \qquad e_{i,d}, f_{i,d}  \mapsto z_{i1}^d \in \CV_{\pm \bs^i}
\end{equation}
which can be easily established by checking the fact that relations \eqref{eqn:rel quad} are respected by the shuffle product \eqref{eqn:shuf prod}. Let us consider the kernel and image of the maps \eqref{eqn:tilde upsilon}
\begin{align}
&K^\pm = \text{Ker } \tUpsilon^\pm \subset \tUUpm \label{eqn:kernel} \\
&\oCS^\pm = \ \text{Im } \tUpsilon^\pm \ \subset \CV^\pm \label{eqn:image}
\end{align} 
The subalgebra $\oCS^+$ will be called the \textbf{(small) shuffle algebra}, to differentiate it from the big shuffle algebra of Definition \ref{def:big shuf}. 

\medskip

\subsection{} 
\label{sub:pairing}

An important role in the present paper will be played by a certain integral pairing, which we will now describe. Let us consider the following notation for all homogeneous rational functions $f(z_1,\dots,z_n)$. If $Dz_a = \frac {dz_a}{2\pi i z_a}$, then we will write
\begin{equation}
\label{eqn:contour integral}
\int_{|z_1| \gg \dots \gg |z_n|} f(z_1,\dots,z_n) \prod_{a=1}^n Dz_a
\end{equation}
for the constant term in the expansion of $f$ as a power series in 
$$
\frac {z_2}{z_1}, \dots, \frac {z_n}{z_{n-1}}
$$
The notation in \eqref{eqn:contour integral} is motivated by the fact that one could compute this constant term as a contour integral (with the contours being concentric circles, situated very far from each other compared to the absolute values of the coefficients of $f$) by invoking the homomorphism $\rho$ in Assumption \ref{assumption}.

\medskip

\begin{definition}
\label{def:pair}

There exists a non-degenerate bilinear pairing \footnote{The reason we employ the notation $\CV^-$ and $\CV^+$ in \eqref{eqn:pair}, despite the fact that the two notations represent identical $\BK$-vector spaces, is the fact that under certain assumptions, \eqref{eqn:pair} can be upgraded to a bialgebra pairing (as in \cite{Arbitrary}).}
\begin{equation}
\label{eqn:pair}
\tUUp \otimes \CV^- \xrightarrow{\langle \cdot, \cdot \rangle} \BK 
\end{equation}
given for all $R \in \CV_{-\bn}$ and all $i_1,\dots,i_n \in I$, $d_1,\dots,d_n \in \BZ$ by
\begin{equation}
\label{eqn:pair formula}
\Big \langle e_{i_1,d_1} \cdots e_{i_n,d_n}, R \Big \rangle = \int_{|z_1| \gg \dots \gg |z_n|} \frac {z_1^{d_1}\dots z_n^{d_n} R(z_1,\dots,z_n)}{\prod_{1\leq a < b \leq n} \zeta_{i_bi_a} \left(\frac {z_b}{z_a} \right)} \prod_{a=1}^n Dz_a
\end{equation}
if $\bs^{i_1}+\dots +\bs^{i_n} = \bn$ and $d_1+\dots+d_n$ equals minus the homogeneous degree of $R$, and 0 otherwise. In the right-hand side of \eqref{eqn:pair formula}, we identify
\begin{equation}
\label{eqn:relabeling}
z_a = z_{i_a\bullet_a}, \quad \forall a \in \{1,\dots, n\}
\end{equation}
where $\bullet_a \in \{1,2,\dots,n_{i_a}\}$ may be chosen arbitrarily due to the symmetry of $R$ (however, we require $\bullet_a \neq \bullet_b$ if $a\neq b$ and $i_a = i_b$). We will call \eqref{eqn:relabeling} a \textbf{relabeling}. 

\end{definition}

\medskip

\noindent There is also an analogous pairing
\begin{equation}
\label{eqn:pair opposite}
\CV^+ \otimes \tUUm \xrightarrow{\langle \cdot, \cdot \rangle} \BK 
\end{equation}
whose formula the interested reader may find in \cite[Definition 2.8]{Arbitrary}. We refer to formulas (2.17), (2.18) and (3.59) of \loccit for the proof of non-degeneracy.

\medskip

\subsection{}

Let $\CS^\mp \subset \CV^\mp$ denote the dual of $K^\pm = \text{Ker }\tUpsilon^\pm$ under the pairings \eqref{eqn:pair} and \eqref{eqn:pair opposite}, respectively, i.e.
\begin{align}
&R^- \in \CS^- \quad \Leftrightarrow \quad \Big \langle K^+, R^- \Big \rangle = 0 \label{eqn:shuf 1} \\
&R^+ \in \CS^+ \quad \Leftrightarrow \quad \Big \langle R^+, K^- \Big \rangle = 0 \label{eqn:shuf 2} 
\end{align}
It is easy to check that $\CS^\pm$ are subalgebras of $\CV^{\pm}$ (in fact, this also follows from the fact that \eqref{eqn:pair} and \eqref{eqn:pair opposite} naturally extend to bialgebra pairings). Thus, we have
$$
\oCS^\pm \subseteq \CS^\pm \qquad \qquad 
$$
because the generators $\{z_{i1}^d\}_{i \in I, d\in \BZ}$ of the algebras on the left lie in the algebras on the right. Moreover, if we consider the \textbf{reduced} quiver quantum toroidal algebra
$$
\UUpm = \tUUpm \Big / K^\pm
$$
then the parings \eqref{eqn:pair} and \eqref{eqn:pair opposite} descend to non-degenerate pairings
\begin{align}
&\UUp \otimes \CS^- \xrightarrow{\langle \cdot, \cdot \rangle} \BK \label{eqn:descended pairing plus} \\
&\CS^+ \otimes \UUm \xrightarrow{\langle \cdot, \cdot \rangle} \BK \label{eqn:descended pairing minus}
\end{align}
One of the main results of \cite{Arbitrary} (specifically, Theorem 1.5 therein) is the following.

\medskip

\begin{theorem} 
\label{thm:equal}

We have $\CS^\pm = \oCS^\pm$, and hence $\tUpsilon^\pm$ induce isomorphisms
\begin{equation}
\label{eqn:iso equal}
\Upsilon^\pm : \UUpm \longrightarrow \CS^\pm
\end{equation}
Moreover, the pairings \eqref{eqn:descended pairing plus} and \eqref{eqn:descended pairing minus} match under these isomorphisms, thus yielding a non-degenerate pairing
\begin{equation}
\label{eqn:descended pairing}
\CS^+ \otimes \CS^- \xrightarrow{\langle \cdot, \cdot \rangle} \BK
\end{equation}

\end{theorem}

\medskip

\noindent We wish to describe $\UUpm$ explicitly, i.e. to give formulas for a system of generators of the kernel $K^\pm$ of the map $\tUUpm \twoheadrightarrow \UUpm$. By formulas \eqref{eqn:shuf 1}--\eqref{eqn:shuf 2}, these sought-for generators are precisely dual to the linear conditions describing the inclusions $\CS^\mp \subset \CV^\mp$. We will exploit this duality in the following Section.

\bigskip

\section{Shuffle algebras for shrubby quivers}
\label{sec:consistent}

\medskip

From now on, we will consider the special case when $Q$ is a quiver drawn on the torus, as in Definition \ref{def:quiver intro}. Moreover, we assume the edges of $Q$ are endowed with parameters $t_e$ as in Subsection \ref{sub:zeta}, and we define the rational functions $\zeta_{ij}(x)$ by formula \eqref{eqn:def zeta}. Our goal is to obtain explicit generators of the ideals $K^\pm$, so that we may realize the reduced quiver quantum toroidal algebras $\UUpm$ as being determined by explicit relations. In what follows, we will only focus on the case $\pm = +$, as the opposite case $\pm = -$ can be obtained by reversing all products.

\medskip

\subsection{}
\label{sub:toric cy}

In Definition \ref{def:relations}, we will construct formal series $e_F$ of elements of $K^+$ associated to the faces of $Q$. When the quiver $Q$ is shrubby (in the sense of Definition \ref{def:consistent intro}), we will show that the coefficients of the series $e_F$ generate $K^+$, thus concluding the proof of Theorem \ref{thm:main}. For every face $F = \{i_0,i_1,\dots,i_{k-1},i_k=i_0\}$ of $Q$, consider
\begin{equation}
\label{eqn:cycle}
t_a = t_{\overrightarrow{i_{a-1}i_a}}
\end{equation}
Note that $t_1\dots t_k = 1$ due to \eqref{eqn:loop constraint}. The arrows in \eqref{eqn:cycle} are the boundary edges of the face $F$ (these edges are uniquely defined, even though it is possible that $Q$ has multiple edges between $i_a$ and $i_b$ for various $ a \neq b$). We will write 
\begin{equation}
\label{eqn:tzeta}
\tzeta_{ij}(x) = \zeta_{ij}(x) (1-x)^{\delta_{ij}} \in \BK[x^{\pm 1}]
\end{equation}
for all $i,j \in I$. For any $1 \leq b \neq c \leq k$, we will write 
\begin{align*}
&\delta_{b<c} = \begin{cases} 1  &\text{if } b < c \\ 0 &\text{if } b > c \end{cases} \\
&\delta_{i_bi_c} = \begin{cases} 1  &\text{if } i_b=i_c \in I \\ 0 &\text{otherwise} \end{cases}
\end{align*}

\medskip

\begin{definition}
\label{def:relations}

For any face $F$ as above, consider the so-called \textbf{clunky relation}
\begin{multline}
\label{eqn:series}
e_F(x_1,\dots,x_k) = \sum_{a=1}^k \frac {x_1t_2\dots t_a}{x_a} \cdot \frac { \prod_{b \succ c} \tzeta_{i_ci_b} \left(\frac {x_c}{x_b} \right)\left( - \frac {x_b}{x_c} \right)^{\delta_{i_bi_c} \delta_{b<c}} }{\prod_{b \sim c + 1} \left(1 - \frac {x_ct_b}{x_b} \right)} \cdot \\ \cdot e_{i_{a}}(x_{a}) \dots e_{i_1}(x_1) e_{i_k}(x_k)  \dots  e_{i_{a+1}}(x_{a+1}) \quad \in \quad  \tUUp[[x_1^{\pm 1}, \dots, x_k^{\pm 1}]]
\end{multline}
In expression \eqref{eqn:series}, the notation $b \succ c$ (respectively $b \sim c+1$) means that $b$ precedes (respectively immediately precedes) $c$ in the sequence $(a,\dots,1,k,\dots,a+1)$.

\end{definition}

\medskip

\begin{proposition}
\label{prop:kernel}

The coefficients of the series \eqref{eqn:series} all lie in $K^+$.

\end{proposition}

\medskip

\begin{proof} Let us consider the formal delta series
$$
\delta\left(z \right) = \sum_{d \in \BZ} z^d
$$
which has the following property for all Laurent polynomials $f(x)$
\begin{equation}
\label{eqn:property delta}
\delta \left( \frac zx \right) f(z) = \delta \left( \frac zx \right) f(x)
\end{equation}
To prove Proposition \ref{prop:kernel}, we must apply the map $\tUpsilon^+$ to the right-hand side of \eqref{eqn:series} and show that the result is 0. By the definition of the shuffle product in \eqref{eqn:shuf prod}, we have
\begin{multline*}
\tUpsilon^+  \left( e_F(x_1,\dots,x_k) \right) = \Sym \left[  \sum_{a=1}^k \frac {x_1t_2\dots t_a}{x_a} \cdot  \right. \\
\left. \frac { \prod_{b \succ c} \tzeta_{i_ci_b} \left(\frac {x_c}{x_b} \right) \prod_{b < c, b \succ c} \left(- \frac {x_b}{x_c} \right)^{\delta_{i_bi_c}} \prod_{c \succ b} \zeta_{i_ci_b} \left(\frac {z_c}{z_b} \right)}{\prod_{b \sim c + 1} \left(1 - \frac {x_ct_b}{x_b} \right)} \cdot \delta\left(\frac {z_1}{x_1} \right) \dots \delta\left(\frac {z_k}{x_k} \right) \right] \stackrel{\eqref{eqn:property delta}}=
\end{multline*}
$$
\Sym \left[  \sum_{a=1}^k \frac {z_1t_2\dots t_a}{z_a} \frac { \prod_{b \succ c} \tzeta_{i_ci_b} \left(\frac {z_c}{z_b} \right) \prod_{b < c, b \succ c} \left(- \frac {z_b}{z_c} \right)^{\delta_{i_bi_c}} \prod_{c \succ b} \zeta_{i_ci_b} \left(\frac {z_c}{z_b} \right)}{\prod_{b \sim c + 1} \left(1 - \frac {z_ct_b}{z_b} \right)} \delta\left(\frac {z_1}{x_1} \right) \dots \delta\left(\frac {z_k}{x_k} \right) \right] 
$$
$$
= \Sym \left[ \sum_{a=1}^k \frac {z_1t_2\dots t_a}{z_a} \frac { \prod_{1 \leq b \neq c \leq k} \zeta_{i_ci_b} \left(\frac {z_c}{z_b} \right) \prod_{b > c, i_b = i_c} \left(1 - \frac {z_c}{z_b} \right)}{\prod_{b \sim c + 1} \left(1 - \frac {z_ct_b}{z_b} \right)} \delta\left(\frac {z_1}{x_1} \right) \dots \delta\left(\frac {z_k}{x_k} \right) \right]
$$
where we let $z_a = z_{i_a\bullet_a}$ as in the relabeling \eqref{eqn:relabeling}, and ``Sym" refers to symmetrization with respect to all $z_a$ and $z_b$ such that $i_a = i_b$.  Therefore, $\tUpsilon^+ (e_F)$ equals
\begin{multline}
\Sym \left[ \frac { \prod_{1 \leq b \neq c \leq k} \zeta_{i_ci_b} \left(\frac {z_c}{z_b} \right) \prod_{b > c, i_b = i_c} \left(1 - \frac {z_c}{z_b} \right)}{\left(1 - \frac {z_1t_2}{z_2} \right)\dots \left(1 - \frac {z_{k-1}t_k}{z_k} \right)\left(1 - \frac {z_kt_1}{z_1} \right) } \cdot \right. \\
 \left. \delta\left(\frac {z_1}{x_1} \right) \dots \delta\left(\frac {z_k}{x_k} \right) \sum_{a=1}^k \frac {z_1t_2\dots t_a}{z_a} \left(1 - \frac {z_a t_{a+1}}{z_{a+1}} \right) \right] \label{eqn:upsilon computation}
\end{multline}
where $z_{k+1}=z_1$. As $t_1\dots t_k  = 1$, the sum in \eqref{eqn:upsilon computation} vanishes, hence so does $\tUpsilon^+ (e_F)$. \end{proof}

\medskip

\subsection{} 

We will now consider the dual to the series $e_F(x_1,\dots,x_k) \in \tUUp[[x_1^{\pm 1}, \dots, x_k^{\pm 1}]]$ under the pairing \eqref{eqn:pair}. We still write $F$ for an arbitrary face of $Q$.

\medskip

\begin{proposition}
\label{prop:vanish}

For any \footnote{The variables of $R$ are relabeled in accordance with \eqref{eqn:relabeling}.} $R(z_1,\dots,z_k) \in \CV_{-\bs^{i_1}-\dots -\bs^{i_k}}$, we have
\begin{equation}
\label{eqn:vanish}
\Big\langle  e_F(x_1,\dots,x_k), R \Big \rangle = 0 \qquad \Leftrightarrow \qquad R\Big|_{z_a = z_{a-1} t_a, \forall a \in \{1,\dots,k\}} = 0
\end{equation}

\end{proposition}

\medskip

\begin{proof} As a consequence of \eqref{eqn:pair formula}, we have
\begin{equation}
\label{eqn:ev}
\Big\langle  e_F(x_1,\dots,x_k), R \Big \rangle = 
\end{equation}
$$
= \sum_{a=1}^k  \ev_{|x_a| \gg \dots \gg |x_1| \gg |x_k| \gg \dots \gg |x_{a+1}|} \left[ \frac {x_1t_2\dots t_a}{x_a} \cdot \frac {R(x_1,\dots,x_k) \prod_{b>c}^{i_b = i_c} \left(1-\frac {x_c}{x_b}\right)  }{\prod_{b \sim c + 1} \left(1 - \frac {x_ct_b}{x_b} \right)} \right] 
$$
where $\ev_{\star}[f]$ denotes the expansion of any rational function $f$ in the region prescribed by the inequalities $\star$. Using the fact that $t_1\dots t_k=1$, it is elementary to prove the following identity of formal series
$$
\sum_{a=1}^k  \ev_{|x_a| \gg \dots \gg |x_1| \gg |x_k| \gg \dots \gg |x_{a+1}|} \left[ \frac {\frac {x_1t_2\dots t_a}{x_a}}{\prod_{b \sim c + 1} \left(1 - \frac {x_ct_b}{x_b} \right)} \right] = \delta\left(\frac {x_1t_2}{x_2} \right) \dots \delta\left( \frac {x_{k-1}t_k}{x_k} \right)
$$
Therefore, the right-hand side of \eqref{eqn:ev} is equal to
$$
\delta\left(\frac {x_1t_2}{x_2} \right) \dots \delta\left( \frac {x_{k-1}t_k}{x_k} \right)  R(x_1,\dots,x_k) \prod_{b>c}^{i_b = i_c} \left(1-\frac {x_c}{x_b}\right)
$$
and vanishes if and only if
\begin{equation}
\label{eqn:vanishes}
R\Big|_{z_a = z_{a-1}t_a, \forall a\in \{1,\dots,k\}} \prod_{b>c}^{i_b = i_c} \left(1-\frac 1{t_{c+1} \dots t_{b}}\right) = 0
\end{equation}
Because of \eqref{eqn:generic 1}, we cannot have $t_{c+1}\dots t_b = 1$ for any $b > c$ with $i_b=i_c$, and therefore \eqref{eqn:vanishes} only holds if $R|_{z_a = z_{a-1}t_a, \forall a\in \{1,\dots,k\}}= 0$, as we needed to show. \end{proof}

\medskip

\noindent More generally, if $R(z_1,\dots,z_n,\dots) \in \CV^-$ is arbitrary, then 
\begin{equation}
\label{eqn:vanish general}
\Big\langle  \tUUp e_F(x_1,\dots,x_k) \tUUp, R \Big \rangle = 0 \qquad \Leftrightarrow \qquad R\Big|_{z_a = z_{a-1}t_a, \forall a \in \{1,\dots,k\}} = 0
\end{equation}
where $z_a$ denotes any variable of $R$ of the form $z_{i_a\bullet_a}$, for all $a \in \{1,\dots,k\}$ (the choice of $\bullet_a$ does not matter due to the symmetry of $R$). Implicit in the notation above is that $R$ may have other variables besides $z_1,\dots,z_n$, and these are not specialized at all. Property \eqref{eqn:vanish general} is proved like \cite[Proposition 3.13]{Loop}; we leave the details as an exercise to the reader. 

\medskip

\subsection{} 

Motivated by Proposition \ref{prop:vanish} and equation \eqref{eqn:vanish general}, we consider the following.

\medskip

\begin{definition}
\label{def:shuffle wheel}

Let $\tCS^\pm \subset \CV^\pm$ denote the subspace consisting of Laurent polynomials $R(z_1,\dots,z_k,\dots)$ such that
\begin{equation}
\label{eqn:wheel}
R\Big|_{z_a = z_{a-1}t_a, \forall a \in \{1,\dots,k\}} = 0
\end{equation}
for any face $F = \{i_0,i_1,\dots,i_{k-1},i_k=i_0\}$ of $Q$ (the notation $t_a$ is that of \eqref{eqn:cycle}).

\end{definition}

\medskip

\noindent We call \eqref{eqn:wheel} a \textbf{wheel condition}, by analogy with the constructions of \cite{E, FO}. It is straightforward to show that $\tCS^\pm$ are closed under the shuffle product, although this will also follow from Proposition \ref{prop:main}. Thus, if we consider the two-sided ideal
$$
J^+ = \Big(\text{series coefficients of }e_F(x_1,\dots,x_k) \Big)_{F \text{ face of }Q} \subset \tUUp
$$
then property \eqref{eqn:vanish general} reads 
\begin{equation}
\label{eqn:iff}
\Big \langle J^+, R \Big \rangle = 0 \qquad \Leftrightarrow \qquad R \in \tCS^-
\end{equation}

\medskip

\begin{remark}
	\label{rem:finite}

Property \eqref{eqn:iff} would still hold if we defined $J^+$ as the ideal generated by a single coefficient of the series $e_F(x_1,\dots,x_k)$ of every possible homogeneous degree in $x_1,\dots,x_k$, for all faces $F$ of the quiver $Q$. In other words, including all the coefficients of all the series $e_F$ as generators of $J^+$ is superfluous; a single coefficient of each homogeneous degree for all faces $F$ would suffice (see \cite[Claim 3.18]{Arbitrary} or \cite[Remark 3.14]{Loop}).

\end{remark}

\medskip

\subsection{}

Proposition \ref{prop:kernel} implies that $J^+ \subseteq K^+$, and therefore
\begin{equation}
\label{eqn:one way}
\tCS^- \supseteq \CS^-
\end{equation}
Our main goal for the remainder of the paper is to prove the opposite inclusion.

\medskip 

\begin{proposition}
\label{prop:main}

If $Q$ is shrubby (as in Definition \ref{def:consistent intro}), then we have 
\begin{equation}
\label{eqn:other way}
\tCS^- \subseteq \CS^-
\end{equation}
and therefore $\tCS^- = \CS^-$. 

\end{proposition}

\medskip

\noindent We also have $\tCS^+ = \CS^+$; the proof is analogous and we will not repeat it.

\medskip

\begin{proof} \emph{of Theorem \ref{thm:main}:} With \eqref{eqn:shuf 1} and \eqref{eqn:iff} in mind, the fact that $\tCS^- = \CS^-$ implies that
$$
\Big\langle K^+, R \Big \rangle = 0 \quad \Leftrightarrow \quad \Big\langle J^+, R \Big \rangle = 0
$$
for any $R \in \CV^-$. If \eqref{eqn:pair} were a pairing of finite-dimensional vector spaces over $\BK$, this would imply that $J^+ = K^+$ and we would be done. In the infinite-dimensional setting at hand, one needs to emulate the proof of \cite[Theorem 1.8]{Arbitrary} to conclude that $J^+ = K^+$. The details are straightforward, and we leave them to the reader. \end{proof}

\medskip

\subsection{}
\label{sub:shrub}

Assume that $Q$ is shrubby, according to Definition \ref{def:consistent intro}, and let $\tQ$ be its universal cover. The following notion will be key to our proof of Proposition \ref{prop:main}.

\medskip

\begin{definition}
\label{def:pre-shrub}

A \textbf{pre-shrub} $S$ is an induced subgraph of $\tQ$ which does not contain the entire boundary of any face, and moreover has the property that if $S$ contains a broken wheel then it must also contain its mirror image. 

\end{definition}

\medskip

\begin{proposition}
\label{prop:no cycles}

A pre-shrub cannot contain any cycles.

\end{proposition}

\medskip

\noindent The Proposition above will be proved in the Appendix. Although a pre-shrub cannot contain any oriented cycles, it can contain unoriented ones (for example, a broken wheel together with its mirror image). The \textbf{interior} of a pre-shrub $S$ is the region completely enclosed by the unoriented cycles belonging to $S$.

\medskip

\noindent Recall that any oriented graph with no cycles yields a partial order on the set of its vertices, with $i>j$ if there exists a path in the graph from $i$ to $j$. Having established that pre-shrubs do not contain any cycles in Proposition \ref{prop:no cycles}, we may consider the corresponding partial order on the set of vertices. With respect to this order, a \textbf{root} of a pre-shrub will refer to a maximal vertex. 

\medskip

\begin{definition}
\label{def:shrub}

A \textbf{shrub} $S$ is a pre-shrub with a single root (thus any vertex of $S$ is reachable from the root by a path), which contains all the vertices in its interior. 

\end{definition}

\medskip 

\noindent An immediate consequence of Definition~\ref{def:consistent intro} is that any two paths in $\tQ$ with the same start and end points can be transformed into one another by the operation of adding and removing cycles, and the operation of replacing broken wheels by their mirror images (the latter operation is called \emph{$F$-term equivalence} in the literature, see for instance \cite{B,D,MR}, following \cite{FHKVW}). Because of \eqref{eqn:loop constraint}, the quantity

\begin{equation}
\label{eqn:product of parameters}
\prod_{e \text{ edge along }p} t_e
\end{equation}
only depends on the start and end points of a path $p$ in $S$. Therefore, for any vertex $i$ in any shrub $S$, we may unambiguously define its \textbf{coordinate}
\begin{equation}
\label{eqn:coordinate}
t_S(i) \in \BK^\times 
\end{equation}
as the product \eqref{eqn:product of parameters} of parameters along any path from the root of $S$ to $i$. The following Proposition will be proved in the Appendix.

\medskip

\begin{proposition}
\label{prop:no edges}

If $i,i'$ are vertices of a shrub $S$, and $i \xrightarrow{e} i'$ is an edge not contained in $S$, then $e$ must be the interface of a broken wheel contained in $S$.

\end{proposition}

\medskip

\subsection{} 
\begin{figure}[H]
\includegraphics[scale=0.40]{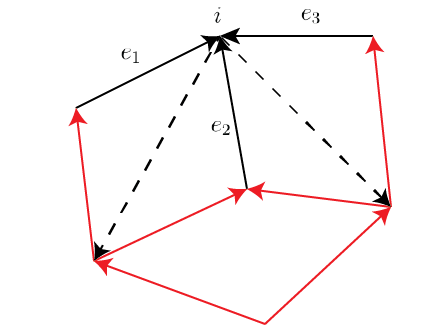} 
\caption{An addable vertex $i$ (in black) to a shrub $S$ (in red).}
\label{fig:3}
\end{figure}

\begin{figure}[H]
\includegraphics[scale=0.40]{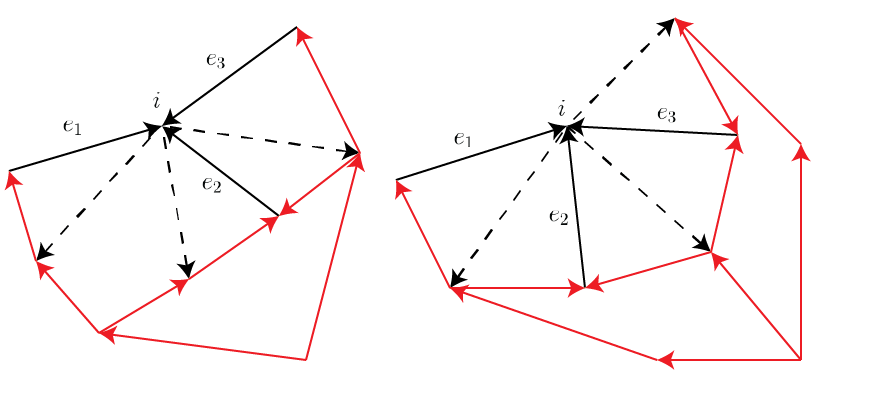} 
\caption{Non-addable vertices $i$ (in black) to shrubs $S$ (in red).}
\label{fig:4}
\end{figure}

Consider a shrub $S \subset \tQ$ and a vertex $i \notin S$. Assume that there are $k>0$ edges from vertices of $S$ to $i$, labeled $e_1,\dots,e_k$ in counterclockwise order around $i$, as in Figures \ref{fig:3} and \ref{fig:4}. The difference between these figures will be explained in Definition \ref{def:addable}, when we discuss the notion of $i$ being addable or non-addable to $S$.

\medskip

\noindent In the situation above, consider any two consecutive edges $e_s$ and $e_{s+1}$ (we make the convention that $e_{k+1} = e_1$). Because $S$ is a shrub (and thus has a root), we may continue these edges in $S$ until they meet, thus yielding paths
\begin{align}
&p_s : j \rightarrow \dots \xrightarrow{e_s} i \label{eqn:path p} \\ 
&p'_s : j \rightarrow \dots \xrightarrow{e_{s+1}} i \label{eqn:path p'}
\end{align}
We may assume the paths $p_s$ and $p_s'$ are simple, non-intersecting (except for the endpoints) and that the region $r_s$ of the plane between $p_s$ and $p_s'$ is minimal with respect to inclusion; this guarantees the uniqueness of $p_s,p_s',r_s$, since the intersection of two minimal regions thus constructed would yield an even smaller acceptable region. Because the vertex $i$ does not belong to the interior of the shrub, a single one of the regions $r_s$ does not contain the counterclockwise angle at $i$ between $e_s$ and $e_{s+1}$. By relabeling the edges if necessary, we assume that the aforementioned region is $r_k$. With this in mind, an index $s\in \{1,\dots,k-1\}$ is called

\medskip

\begin{itemize}[leftmargin=*]

\item \textbf{good} if $p_s$ and $p'_s$ are broken wheels, which are mirror images of each other

\medskip

\item \textbf{bad} if there exist edges $i \xrightarrow{e} v \in p_s$ and $i \xrightarrow{e'} v' \in p'_s$ with $v,v' \neq j$, such that the sub-regions of $r_s$ between $e$ and $p_s$ (respectively between $e'$ and $p'_s$) are faces

\end{itemize}

\medskip

\noindent For example, both $s \in \{1,2\}$ in Figure \ref{fig:3} are good. However, in the picture on the left of Figure \ref{fig:4}, $s = 1$ is bad and $s = 2$ is good. Meanwhile, we call the index $s = k$ 

\medskip

\begin{itemize}[leftmargin=*]

\item \textbf{good} if there are no edges from $i$ to $S$ in the counterclockwise region from $e_k$ to $e_1$ (i.e. the region $\BR^2 \backslash r_k$); this is the case in Figure \ref{fig:3}.

\medskip

\item \textbf{bad} if there exists an edge from $i$ to $S$ in the region $\BR^2 \backslash r_k$, which determines a face together with the other edges in $S$ and exactly one of the edges $e_1$ and $e_k$; this is the case in the picture on the right of Figure \ref{fig:4}.

\medskip

\end{itemize}

\noindent The following result will be proved in the Appendix. 

\medskip

\begin{proposition}
\label{prop:good or bad}

For $i \notin S$ as above, every $s \in \{1,\dots,k\}$ is either good or bad.

\end{proposition}

\medskip

\subsection{}
\label{sub:broken wheels}

If $S$ is a shrub and $i \notin S$, let $S+i$ denote the subgraph obtained from $S$ by adding the vertex $i$ and the edges from $S$ to $i$ (we assume such edges exist).

\medskip

\begin{definition}
\label{def:addable}

In the situation above, we call $i$ \textbf{addable} to $S$ if all $s \in \{1,\dots,k\}$ are good, and \textbf{non-addable} to $S$ otherwise. 

\end{definition}

\medskip

\noindent Figures \ref{fig:3} and \ref{fig:4} provide examples of addable and non-addable vertices. This terminology is motivated by the following result, which will be proved in the Appendix. 

\medskip

\begin{proposition}
\label{prop:key}

Assume $S \subset \tQ$ is a shrub and $i \notin S$ is a vertex. Then $S+i$ is a shrub if and only if $i$ is an addable vertex to $S$.

\end{proposition}

\medskip

\noindent The main distinction to us between addable and non-addable vertices is the following result, which will also be proved in the Appendix.

\medskip

\begin{proposition}
\label{prop:count}

Assume $S \subset \tQ$ is a shrub and $i \notin S$ is a vertex with $k > 0$ edges from $S$ to $i$. The maximal number of broken wheels in $S+i$ that all pass through $i$ and do not pairwise intersect at any other vertex is
$$
\begin{cases} k-1 &\text{if }i\text{ is addable to }S \\ \geq k &\text{otherwise} \end{cases}
$$

\end{proposition}

\medskip

\subsection{} We are now ready to prove Proposition \ref{prop:main}. Fix positive real numbers
\begin{equation}
\label{eqn:rho}
\{ \rho_i \}_{i \in I} \subset \BR_{>0}
\end{equation}
and extend the notation $\rho_i$ to $i \in \tQ$ via the projection $\pi : \tQ \rightarrow Q$. We choose the positive real numbers \eqref{eqn:rho} to be generic enough so that for any pair of paths
$$
i \xrightarrow{e_1} \dots \xrightarrow{e_k} j \xleftarrow{e_\ell'} \dots \xleftarrow{e_1'} i'
$$
in $\tQ$ for which $\pi(i) \neq \pi(i') \in Q$, we have
\begin{equation}
\label{eqn:generic 2}
\rho_i |t_{e_1}| \dots |t_{e_k}| \neq \rho_{i'} |t_{e'_1}| \dots |t_{e'_{\ell}}| 
\end{equation} 
(recall that $|t_e|$ was defined in \eqref{eqn:abs}). Because of \eqref{eqn:generic 1} and the fact that $\pi : \tQ \rightarrow Q$ is a covering, formula \eqref{eqn:generic 2} actually holds for all $i \neq i' \in \tQ$. Define (cf.~\eqref{eqn:coordinate})
\begin{equation}
\label{eqn:positive coordinate}
|t_S(i)| = |t_{e_1}| \dots |t_{e_k}| \in \BR_{>0}
\end{equation} 
for any vertex $i$ in any shrub $S$, where $e_1,\dots,e_k$ is any path from the root to $i$.

\medskip

\begin{definition} 
\label{def:acceptable}

A shrub $S$ is called \textbf{acceptable} if the numbers \eqref{eqn:positive coordinate} satisfy
\begin{equation}
\label{eqn:acceptable} 
|t_S(i)| > \frac {\rho_i}{\rho_j}
\end{equation}
for any vertex $i$ of $S$ other than the root $j$ of $S$.

\end{definition} 

\medskip 

\noindent In the subsequent proof, we will identify shrubs up to translations with respect to the fibers of the covering $\pi : \tQ \rightarrow Q$. This can be visualized by fixing a right inverse $s$ of $\pi$ and restricting attention to shrubs rooted at vertices in the image of $s$.

\medskip 

\begin{proof} \emph{of Proposition \ref{prop:main}:} Let us consider any
$$
\phi = \mathop{\sum_{i_1,\dots,i_n \in I}}_{d_1,\dots,d_n \in \BZ} \text{coefficient} \cdot e_{i_1,d_1} \dots e_{i_n,d_n} \in K^+
$$
and any $R \in \tCS^-$. Our goal is to show that
\begin{equation}
\label{eqn:goal}
\Big \langle \phi, R \Big \rangle = 0
\end{equation}
as this would imply that $R \in \CS^-$, as required. Recall from formula \eqref{eqn:pair formula} that
\begin{equation}
\label{eqn:pairing final}
\Big \langle e_{i_1,d_1} \cdots e_{i_n,d_n}, R \Big \rangle = \int_{|z_1| \gg \dots \gg |z_n|} f(z_1,\dots,z_n) \prod_{a=1}^n Dz_a
\end{equation}
where in the notation \eqref{eqn:relabeling}, we have
\begin{equation}
\label{eqn:function}
f(z_1,\dots,z_n) = \frac {z_1^{d_1}\dots z_n^{d_n} R(z_1,\dots,z_n)}{\prod_{1\leq a < b \leq n} \zeta_{i_bi_a} \left(\frac {z_b}{z_a} \right)}
\end{equation}
A \textbf{labeling} of a shrub $S \subset \tQ$ will refer to a totally ordered labeling of its vertices
\begin{equation}
\label{eqn:ordered vertices}
\{ i_{a_1},\dots,i_{a_s} \} \subset \tQ
\end{equation}
(for certain $1\leq a_1<\dots<a_s \leq n$) such that if there is an arrow $i_{x} \rightarrow i_y$ between vertices of $S$, then $x>y$. In particular, the root of $S$ must be the vertex $i_{a_s}$.

\medskip

\begin{proposition}
\label{prop:residue}

For any labeled shrub $S$ and function $f$ as in \eqref{eqn:function}, define
\begin{equation}
\label{eqn:residue}
\mathop{\emph{Res}}_S f \in \BK(\{z_a\}_{a \in \{1,\dots,n\} \backslash \{a_1,\dots,a_{s-1}\}})
\end{equation}
by the following recursive procedure: 

\medskip 

\begin{itemize}[leftmargin=*]

\item input $f_s = f$;

\medskip 

\item for all $x$ from $s-1$ down to $1$, we claim that $f_{x+1}$ has at most a simple pole at
\begin{equation}
\label{eqn:pole}
z_{a_{x}} = z_{a_s} t_S(i_{a_{x}})
\end{equation}
Define $f_x$ to be the residue at the above pole;

\medskip 

\item output $\mathop{\emph{Res}}_S f = f_1$.

\end{itemize}

\medskip 

\noindent Moreover, the quantity $\mathop{\emph{Res}}_S f$ only depends on $S$ and not on the labeling \eqref{eqn:ordered vertices}, as long as we have $x>y$ any time there is an edge $i_x \rightarrow i_y$ between vertices of $S$.

\end{proposition}

\medskip 

\begin{proof} We must first prove the claim on the simplicity of the pole at \eqref{eqn:pole}. To this end, fix $x \in \{1,\dots,s-1\}$ and consider the induced subgraph $S' \subset S$ consisting of all vertices $i_{a_y}$ with $y > x$. It is easy to see that $S'$ is a shrub and that $i_{a_x}$ is an addable vertex to $S'$. Therefore, we may assume that there are $k > 0$ edges
$$
i_{b_1} \xrightarrow{e_1} i_{a_x}, \dots, i_{b_k} \xrightarrow{e_k} i_{a_x}
$$
from the shrub $S'$ to the vertex $i_{a_x}$, for certain $b_1,\dots,b_k > a_x$. Since these edges must be distributed as in Figure \ref{fig:3}, the denominator of \eqref{eqn:function} includes the $k$ factors
$$
1 - \frac {z_{b_1} t_{e_1}}{z_{a_{x}}}, \dots, 1 - \frac {z_{b_k} t_{e_k}}{z_{a_{x}}}
$$
Once the variables $z_{b_1},\dots,z_{b_k}$ are specialized to $z_{a_s}t_S(i_{b_1}),\dots, z_{a_s}t_S(i_{b_k})$, respectively, the fact that $t_S(i_{a_x}) = t_S(i_{b_1})t_{e_1} = \dots = t_S(i_{b_k})t_{e_k}$ implies that the denominator of \eqref{eqn:function} will feature the factor
$$
\left(1 - \frac {z_{a_s} t_S(i_{a_x})}{z_{a_{x}}} \right)^k
$$
Thus, to prove that the pole invoked in the statement of the Proposition is at most simple, we need to show that the numerator of \eqref{eqn:function} vanishes to order at least $k-1$ at the specialization \eqref{eqn:pole}. However, the numerator of $f$ vanishes whenever any subset of its variables are specialized according to \eqref{eqn:wheel} for any face $F$. As there exist $k-1$ broken wheels whose only common vertex is $i_{a_x}$ (see Proposition \ref{prop:count}), property \eqref{eqn:wheel} for the $k-1$ faces enclosed by said broken wheels implies that the numerator of $f$ vanishes to order $\geq k-1$ at the specialization \eqref{eqn:pole}. \footnote{In claiming the vanishing of the numerator of $f$ to order at least $k-1$, we are invoking the fact that for any $k, \ell_1,\dots, \ell_{k-1} \in \BN$, we have
$$
\bigcap_{c=1}^{k-1} \left(x_c^{(1)},\dots,x_c^{(\ell_c)}\right) = \left(x_1^{(\alpha_1)} x_2^{(\alpha_2)}\dots x_{k-1}^{(\alpha_{k-1})} \right)_{\alpha_1 \in \{1,\dots,\ell_1\},\dots, \alpha_{k-1} \in \{1,\dots,\ell_{k-1}\}}
$$
in the ring of polynomials over \underline{distinct} variables $\{x_c^{(1)},\dots,x_c^{(\ell_c)}\}_{c \in \{1,\dots,k-1\}}$.}

\medskip 

\noindent To see that the quantity \eqref{eqn:residue} is independent on the labeling and only depends on the underlying shrub $S$, we argue as follows. $\mathop{\text{Res}}_S f$ can be interpreted as a multivariate integral over the following nested contours:
$$
z_{a_{s-1}} \text{ goes around } z_{a_s} t_S(i_{a_{s-1}}), \dots,  z_{a_{1}} \text{ goes around } z_{a_s} t_S(i_{a_1})
$$
The aforementioned independence is merely saying that we can swap the order of the contours of $z_{a_x}$ and $z_{a_y}$ if there are no edges between vertices $i_{a_x}$ and $i_{a_y}$. However, the non-existence of such edges precludes any linear factors of the form
$$
1 - \frac {z_{a_x}t_e}{z_{a_y}} \quad \text{or} \quad 1 - \frac {z_{a_y}t_e}{z_{a_x}}
$$
in the denominator of \eqref{eqn:function}, which allows us to swap the order of the contours. \end{proof}

\medskip 

\noindent An $m$-\textbf{labeled shrubbery} $\mathscr{S}$ is a disjoint union of labeled shrubs in $\tQ$, whose $n-m+1$ vertices are endowed with distinct labels among the numbers $m, m+1, \dots, n$. Note that we do allow the different shrubs in a shrubbery to overlap, e.g.
$$
\{i_m\} \sqcup \dots \sqcup \{i_n\} 
$$
is an $m$-labeled shrubbery consisting of one-vertex shrubs, which is well-defined even if $i_x = i_y$ for certain $x \neq y$. An $m$-labeled shrubbery is called \textbf{acceptable} if all of its constituent shrubs are acceptable, with respect to the numbers \eqref{eqn:rho}.

\medskip

\begin{claim}
\label{claim:final}

For any $m \in \{1,\dots,n\}$, consider
\begin{multline}
X_m = \sum^{m\text{-labeled acceptable}}_{\text{shrubberies }\mathscr{S} = S_1 \sqcup \dots \sqcup S_t} \int_{|z_1| \gg \dots \gg |z_{m-1}|\gg |z_{r_1}| = \rho_{i_{r_1}}, \dots, |z_{r_t}| = \rho_{i_{r_t}}} \\ \mathop{\emph{Res}}_{S_1} \dots \mathop{\emph{Res}}_{S_t} f \prod_{a=1}^{m-1} Dz_a \prod_{u = 1}^t Dz_{r_u} \label{eqn:multline}
\end{multline}
where $i_{r_1},\dots,i_{r_t}$ are the roots of the shrubs $S_1,\dots,S_t$. Then we have 
\begin{equation}
\label{eqn:recursion}
X_{m-1} = X_m
\end{equation}
for all $m \in \{2,\dots,n\}$.

\end{claim}

\medskip

\noindent Note that there are finitely many $m$-labeled shrubberies, due to the fact that shrubs that only differ by a translation of $\tQ$ over $Q$ are identified (more specifically, right before the proof of Proposition~\ref{prop:main} we prescribed the roots of all our shrubs to lie in a particular set of preimages of $\pi : \tQ \rightarrow Q$). Consider any $u \neq v$ in $\{1,\dots,t\}$; formula \eqref{eqn:function} implies that the integrand of \eqref{eqn:multline} has linear factors of the form
$$
1 - \frac {z_{r_u} t_S(i_x) t_e}{z_{r_v} t_S(i_y)}
$$
in the denominator, where $e$ is any edge from any vertex $i_x$ in the shrub $S_u$ to any vertex $i_y$ in the shrub $S_v$. Because of \eqref{eqn:generic 2} and the sentence following it, the linear factors above do not create any poles involving the variables $z_{r_u}$ and $z_{r_v}$ on the circles $|z_{r_u}| = \rho_{r_u}$, $|z_{r_v}| = \rho_{r_v}$, unless $r_u = r_v$ and the pole is at $z_{r_u} = z_{r_v}$. We will show at the very end of the proof of Proposition~\ref{prop:main} that the latter pole actually vanishes in \eqref{eqn:multline}, thus ensuring that $X_m$ is well-defined as a contour integral.

\medskip

\begin{proof} \emph{of Claim~\ref{claim:final}:} To prove \eqref{eqn:recursion}, one needs to move the contour of the variable $z_{m-1}$ from $\infty$ toward the circle of radius $\rho_{i_{m-1}}$. If the contour reaches the circle, this corresponds to adding the one-vertex shrub $\{i_{m-1}\}$ to the shrubbery $\mathscr{S}$. Otherwise, the variable $z_{m-1}$ must be ``caught" in one of the poles of the form
\begin{equation}
\label{eqn:factor}
1 - \frac {z_b t_{e}}{z_{m-1}}
\end{equation}
for some $b > m-1$ and some edge $e = \overrightarrow{i_b i_{m-1}}$. Assume $i_b$ belongs to one of the constituent shrubs $S_u \subset \mathscr{S}$, and suppose there is a number $k > 0$ of edges from the shrub $S_u$ to $i_{m-1}$. Then we have one of the following three possibilities.

\medskip

\begin{itemize}[leftmargin=*]

\item If the vertex $i_{m-1}$ is addable to $S_u$ as in Definition \ref{def:addable}, then Proposition \ref{prop:key} implies that $S_u' = S_u+i_{m-1}$ is a shrub. Thus, the operation
\begin{multline*}
m\text{-labeled shrubbery } \mathscr{S} = S_1 \sqcup \dots \sqcup S_u \sqcup \dots \sqcup S_t \leadsto \\ \leadsto (m-1)\text{-labeled shrubbery } \mathscr{S}' = S_1 \sqcup \dots \sqcup S'_u \sqcup \dots \sqcup S_t
\end{multline*}
shows how to obtain $X_{m-1}$ by applying the contour moving procedure to $X_m$ (the fact that we only encounter acceptable shrubs is due to the fact that we move the contour of $z_{m-1}$ from $\infty$ to the circle of radius $\rho_{i_{m-1}}$, but no further). 

\medskip

\item If the vertex $i_{m-1}$ is non-addable to $S_u$, then Proposition \ref{prop:count} states that there exist $k$ broken wheels completely contained in $S_u + i_{m-1}$ that only intersect pairwise at the vertex $i_{m-1}$. As we have seen at the end of the proof of Proposition \ref{prop:residue}, this means that the numerator of $f$ has enough factors to cancel the $k$ copies of the factor \eqref{eqn:factor} from the denominator of $f$. We conclude that non-addable vertices do not produce actual poles in the integral \eqref{eqn:multline}. 

\medskip

\item If the vertex $i_{m-1}$ is already in $S_u$ (suppose that it corresponds to a variable $z_c$ for some $c > m-1$), then the linear factor of $z_{m-1}-z_c$ in the denominator of 
$$
\zeta_{i_ci_{m-1}} \left(\frac {z_c}{z_{m-1}}\right)
$$
allows the numerator of $f$ to annihilate the pole of the form \eqref{eqn:factor}. \end{itemize} \end{proof}

\medskip

\noindent Repeated applications of Claim \ref{claim:final} imply that $X_1 = X_n$. Since $X_n$ is tautologically the same as the right-hand side of \eqref{eqn:pairing final}, we conclude that
\begin{multline}
\Big \langle e_{i_1,d_1} \cdots e_{i_n,d_n}, R \Big \rangle  = \sum^{1\text{-labeled acceptable}}_{\text{shrubberies }\mathscr{S} = S_1 \sqcup \dots \sqcup S_t} \\ \int_{|z_{r_1}| = \rho_{i_{r_1}}, \dots, |z_{r_t}| = \rho_{i_{r_t}}} \mathop{\text{Res}}_{S_1} \dots \mathop{\text{Res}}_{S_t} \frac {z_1^{d_1}\dots z_n^{d_n} R(z_1,\dots,z_n)}{\prod_{1\leq a < b \leq n} \zeta_{i_bi_a} \left(\frac {z_b}{z_a} \right)} \prod_{u = 1}^t Dz_{r_u} \label{eqn:pairing shrubberies}
\end{multline}
The fact that the number $\rho_i$ only depends on the image of $i$ in $Q$ means that we can color-symmetrize the integrand (i.e. symmetrize it with respect to those variables $z_a$ and $z_b$ such that $i_a = i_b$) without changing the value of the integral: 
\begin{multline*}
\Big \langle e_{i_1,d_1} \cdots e_{i_n,d_n}, R \Big \rangle  = \sum^{\text{fixed }1\text{-labeled acceptable}}_{\text{shrubberies } \bar{\mathscr{S}} = \bar{S}_1 \sqcup \dots \sqcup \bar{S}_t} \\ \int_{|z_{r_1}| = \rho_{i_{r_1}}, \dots, |z_{r_t}| = \rho_{i_{r_t}}} \mathop{\text{Res}}_{\bar{S}_1} \dots \mathop{\text{Res}}_{\bar{S}_t} \text{ Sym} \left[ \frac {z_1^{d_1}\dots z_n^{d_n} R(z_1,\dots,z_n)}{\prod_{1\leq a < b \leq n} \zeta_{i_bi_a} \left(\frac {z_b}{z_a} \right)} \right] \prod_{u = 1}^t Dz_{r_u} \end{multline*}
where the adjective ``fixed" means that we are summing over a single 1-labeled acceptable shrubbery in every equivalence class given by permuting those $a \neq b$ such that $i_a = i_b$. The previous statement uses the fact that the rational function \eqref{eqn:function} has a non-zero residue at a certain labeled shrub $S$ only if the arrows of $S$ point from bigger labels to smaller labels. Because of the identity
$$
\tUpsilon^+(e_{i_1,d_1} \cdots e_{i_n,d_n}) \stackrel{\eqref{eqn:shuf prod}}= \Sym \left[ z_1^{d_1}\dots z_n^{d_n}  \prod_{1\leq a < b \leq n} \zeta_{i_ai_b} \left(\frac {z_a}{z_b} \right) \right] 
$$
we obtain
\begin{equation}
\label{eqn:pairing sym shrubberies}
\Big \langle e_{i_1,d_1} \cdots e_{i_n,d_n}, R \Big \rangle  = \sum^{\text{fixed }1\text{-labeled acceptable}}_{\text{shrubberies } \bar{\mathscr{S}} = \bar{S}_1 \sqcup \dots \sqcup \bar{S}_t} 
\end{equation}
$$
\int_{|z_{r_1}| = \rho_{i_{r_1}}, \dots, |z_{r_t}| = \rho_{i_{r_t}}} \mathop{\text{Res}}_{\bar{S}_1} \dots \mathop{\text{Res}}_{\bar{S}_t} \frac {\tUpsilon^+(e_{i_1,d_1} \cdots e_{i_n,d_n}) R(z_1,\dots,z_n)}{\prod_{1\leq a \neq b \leq n} \zeta_{i_bi_a} \left(\frac {z_b}{z_a} \right)} \prod_{u = 1}^t Dz_{r_u} 
$$
We conclude that $\langle \phi, R \rangle$ is a linear functional of $\tUpsilon^+(\phi)$. Since the latter expression is 0 due to the fact that $\phi \in K^+$, we infer the required formula \eqref{eqn:goal}.

\medskip 

\noindent In order to establish the well-defined-ness of $X_m$, it remains to show that the RHS of \eqref{eqn:multline} does not have poles at $z_{r_u} = z_{r_v}$ if $r_u = r_v$ and $u \neq v$. To keep the explanation simple, we will exclude from the notation those variables which do not correspond to the shrubs $S_u$ and $S_v$. Thus, we need to prove that the quantity
\begin{equation}
\label{eqn:ym}
Y_m = \int_{|z_1| \gg \dots \gg |z_{m-1}|\gg |z_{r_u}| = |z_{r_v}|} \sum^{m\text{-labeled acceptable}}_{\text{shrubberies }\mathscr{S} = S_u \sqcup S_v} \mathop{\text{Res}}_{S_u} \mathop{\text{Res}}_{S_v} f \prod_{a=1}^{m-1} Dz_a
\end{equation}
does not have poles at $z_{r_u} = z_{r_v}$, and we will proceed by descending induction on $m$. \footnote{In \eqref{eqn:ym}, we moved the integral outside of the summation, cf.~\eqref{eqn:multline}. This is because the individual terms of the summation do have poles at $z_{r_u} = z_{r_v}$, but the subsequent argument will establish that the entire sum does not.} To establish the induction step, we use \eqref{eqn:function} to write (let $t(i) = t_{S_u}(i) = t_{S_v}(i)$)
\begin{equation}
\label{eqn:double residue}
\mathop{\text{Res}}_{S_u} \mathop{\text{Res}}_{S_v} f = \frac {g(z_{m-1})}{\prod_{i \in \tQ} \left[ \left(1 - \frac {z_{r_u}t(i)}{z_{m-1}}\right)^{k_i - \chi(i \in S_u)} \left(1 - \frac {z_{r_v}t(i)}{z_{m-1}}\right)^{\ell_i - \chi(i \in S_v)} \right]}
\end{equation}
where $g(z_{m-1})$ is a Laurent polynomial (which may have denominators in $z_1,\dots,z_{m-2}$ and $z_{r_u},z_{r_v}$ other than $z_{r_u}-z_{r_v}$, but this will not affect the proof), and we let
\medskip 

\begin{itemize} 

\item $k_i$ be the number of edges from vertices of the shrub $S_u$ to any given $i \in \tQ$;

\medskip 

\item $\ell_i$ be the number of edges from vertices of the shrub $S_v$ to any given $i \in \tQ$;

\medskip 

\item $\chi(P)$ be $1$ or $0$ depending on whether the statement $P$ is true or false.

\end{itemize} 

\medskip 

\noindent Arguing just like in the proof of Claim \ref{claim:final}, we use the wheel conditions \eqref{eqn:wheel} and Proposition~\ref{prop:count} to infer that there exists a Laurent polynomial $h(z_{m-1})$ (which may have denominators in $z_1,\dots,z_{m-2}$ and $z_{r_u},z_{r_v}$ other than $z_{r_u}-z_{r_v}$) such that
$$
g(z_{m-1}) = h(z_{m-1}) \prod_{i \in \tQ} 
$$
$$ 
\left[ \left( 1 - \frac {z_{r_u}t(i)}{z_{m-1}} \right)^{k_i-\chi(i \in S_u) - \chi(i \text{ addable to }S_u)}
\left(1 - \frac {z_{r_v}t(i)}{z_{m-1}} \right)^{\ell_i-\chi(i\in S_v) - \chi(i \text{ addable to }S_v)} \right]
$$
Because of this, formula \eqref{eqn:double residue} yields the following residue computation 
\begin{align*}
&\sum_{i \in \tQ} \left( \mathop{\text{Res}}_{z_{m-1} = z_{r_u}t(i)} + \mathop{\text{Res}}_{z_{m-1} = z_{r_v}t(i)}  \right)\mathop{\text{Res}}_{S_u} \mathop{\text{Res}}_{S_v} f = \\
&\sum_{i \in \tQ} \left( \mathop{\text{Res}}_{z_{m-1} = z_{r_u}t(i)} + \mathop{\text{Res}}_{z_{m-1} = z_{r_v}t(i)}  \right) \frac {h(z_{m-1})}{\prod_{i \in \tQ} \left[ \left(1 - \frac {z_{r_u}t(i)}{z_{m-1}}\right)^{\chi(i \text{ addable to } S_u)} \left(1 - \frac {z_{r_v}t(i)}{z_{m-1}}\right)^{\chi(i \text{ addable to } S_v)} \right]} \\
&=\sum_{i \text{ addable to }S_u \text{ but not to }S_v} h(z_{r_u}t(i)) + \sum_{i \text{ addable to }S_v \text{ but not to }S_u} h(z_{r_v}t(i)) \\
&+\sum_{i \text{ addable to both }S_u \text{ and }S_v} \left( \frac {h(z_{r_u}t(i))}{1 - \frac {z_{r_v}}{z_{r_u}}} + \frac {h(z_{r_v}t(i))}{1 - \frac {z_{r_u}}{z_{r_v}}} \right) 
\end{align*}
Since the pole at $z_{r_u} = z_{r_v}$ cancels out in the right-hand side of the expression above, the induction step is complete, and the quantity \eqref{eqn:ym} is well-defined. \end{proof}

\medskip

\noindent Note that \eqref{eqn:pairing sym shrubberies} implies the following formula for the descended pairing \eqref{eqn:descended pairing}:
\begin{equation}
\label{eqn:formula pairing shrubberies}
\Big \langle R^+, R^- \Big \rangle  = \sum^{\text{fixed }1\text{-labeled acceptable}}_{\text{shrubberies } \bar{\mathscr{S}} = \bar{S}_1 \sqcup \dots \sqcup \bar{S}_t} 
\end{equation}
$$
\int_{|z_{r_1}| = \rho_{i_{r_1}}, \dots, |z_{r_t}| = \rho_{i_{r_t}}} \mathop{\text{Res}}_{\bar{S}_1} \dots \mathop{\text{Res}}_{\bar{S}_t} \text{ Sym} \left[ \frac {R^+(z_1,\dots,z_n)R^-(z_1,\dots,z_n)}{\prod_{1\leq a \neq b \leq n} \zeta_{i_bi_a} \left(\frac {z_b}{z_a} \right)} \right] \prod_{u = 1}^t Dz_{r_u} 
$$
for any $R^\pm \in \tCS^\pm$ of opposite degrees. Formula \eqref{eqn:formula pairing shrubberies} shows that shrubberies are not just technical tools used in the proof of Proposition \ref{prop:main}, but natural combinatorial objects which parameterize the summands in the formula for the pairing \eqref{eqn:descended pairing}.

\bigskip

\section{Appendix: the joys of gardening}
\label{sec:gardening}

\medskip

\noindent In the present Section, we will motivate our notion of shrubby quivers by relating it with more traditional consistency conditions in the theory of brane tilings and dimer models. We also prove several technical results from Section \ref{sec:consistent}.

\medskip

\subsection{}

\noindent Let $Q$ denote a quiver in $\BT^2$, as in Definition \ref{def:quiver intro}, i.e. the faces of $Q$ are colored in blue/red such that any two faces which share an edge have different colors. 

\medskip

\begin{definition}
\label{def:r charge}

A non-degenerate $R$-charge (see for instance \cite{HHV, HV}) is a function
$$
R : E \rightarrow (0,1)
$$
such that for any vertex $i$ and any face $F$ of the quiver $Q$, we have
\begin{align*}
&\sum_{e \text{ edge around }F} R(e) = 2 \\
&\sum_{e \text{ edge incident to }i} (1-R(e)) = 2
\end{align*}
\footnote{Loops at $i$ are counted twice in the formula above.} Geometrically, the properties above imply that the quiver $Q$ can be drawn on the torus so that all faces are polygons circumscribed in circles of the same radius, and the centers of these circles lie strictly inside the faces (the number $\pi R(e)$ is the central angle subtended by the chord $e$ in the aforementioned circles). 

\end{definition}

\medskip

\begin{figure}[h]
\includegraphics[scale=0.55]{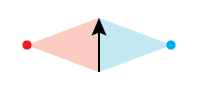} 
\caption{A rhombus. The blue/red bullets represent the centers of the blue/red faces, while the other two vertices of the rhombus are vertices of $Q$ (with an arrow between them).}
\label{fig:5}
\end{figure}

\noindent The existence of a non-degenerate $R$-charge allows one to define a rhombus tiling of the torus, as follows. Draw the centers of the (circles circumscribing the) blue/red polygonal faces as blue/red bullets. Then the condition that the segments between the vertices and the bullets all have the same length means that $\BT^2$ is tiled by rhombi. To recover the arrows in the quiver $Q$ from the rhombus tiling, one need only draw the diagonals between non-bullet vertices of the rhombi, and orient them so that they keep the blue/red bullets on the right/left (see Figure \ref{fig:5}).

\medskip

\medskip

\noindent Recall the notion of shrubby quivers from Definition \ref{def:consistent intro}. Lemma \ref{lem:non-degenerate is shrubby} below is proved just like \cite[Lemma 5.3.1]{HHV} (note that the topology of shrubby quivers underlies the notion of $F$-term equivalent paths, see \cite[Definition 2.5]{D} and \cite[Condition 4.12]{MR}).

\medskip

\begin{lemma}
\label{lem:non-degenerate is shrubby}

If there exists a non-degenerate $R$-charge, then $Q$ is shrubby.

\end{lemma}

\medskip

\subsection{} 
\label{sub:broken}

In the remainder of the paper, we provide proofs of some technical results about shrubs and pre-shrubs, specifically Propositions \ref{prop:no cycles}, \ref{prop:no edges}, \ref{prop:good or bad}, \ref{prop:key} and \ref{prop:count}. Throughout the present Section, we assume $Q$ to be a shrubby quiver, with universal cover $\tQ$. All paths and cycles in a quiver are understood to be oriented.

\medskip

\begin{definition}
\label{def:region and area}

Given two paths $p$ and $p'$ in $\tQ$ with the same start and end points, we will write $r(p,p')$ for the closed \textbf{region} inside $\BR^2$ contained between $p$ and $p'$. The \textbf{area} of this region, denoted by $a(p,p') \in \BN$, will refer to the number of faces contained inside $r(p,p')$. In particular, if $C$ is a cycle, we will write $r(C)$ and $a(C)$ for the closed region and area (respectively) contained inside $C$.

\end{definition}

\medskip

\begin{proof} \emph{of Proposition \ref{prop:no cycles}:} Assume for the purpose of contradiction that a pre-shrub $S$ contains a cycle, and let us fix such a cycle $C$ of minimal area (as in Definition \ref{def:region and area}). We must have $a(C) > 2$, since otherwise $C$ would be the boundary of a face, or the union of boundaries of two faces which meet at a single point, both situations being forbidden for pre-shrubs. Definition \ref{def:consistent intro} for $p=C$ and $p' = \text{trivial}$ implies that there exist two adjacent faces (as in Figure \ref{fig:2}) for which e.g. the red path is completely contained in $C$, and the red and blue regions are contained inside $r(C)$. By the defining property of a pre-shrub, $S$ also contains the blue path. Thus, the cycle
$$
C' = C - \{\text{red path}\} + \{\text{blue path}\}
$$
is contained in $S$, and moreover $a(C') = a(C) - 2$. This contradicts the minimality of the area of $C$. \end{proof}

\medskip

\begin{proof} \emph{of Proposition \ref{prop:no edges}:} Assume that $e$ is an edge from vertex $i$ to vertex $i'$, where $i,i' \in S$ but $e \not\subset S$. By the very definition of the root $r$ of a shrub, there are paths from $r$ to $i$ and $i'$, respectively. Following the aforementioned paths until they first intersect, we conclude that there exist simple paths
\begin{align*}
&p : j \rightarrow \dots \rightarrow i \\ 
&p' : j \rightarrow \dots \rightarrow i' 
\end{align*}
with no vertices in common other than the source $j$. We have three scenarios.

\medskip

\noindent (1) If $j=i$, then $e$ and $p'$ are both paths from $i$ to $i'$. We may assume that $p'$ is chosen such that $a(e,p')$ is minimal. Definition \ref{def:consistent intro} implies that $p'$ contains a broken wheel $B$ (since $e$ consists of a single edge, it cannot contain a broken wheel). Since $S$ is a shrub, it therefore contains the mirror image $B'$ of $B$. Thus, if we modify $p'$ by replacing its sub-path $B$ with $B'$, then we contradict the minimality of $a(e,p')$. We conclude that this scenario is impossible.

\medskip

\noindent (2) If $j = i'$, then $C = p \cup e$ is a cycle, and we assume that $p$ is chosen so that $a(C)$ is minimal. If $a (C) = 1$ then we are done (since $r(C)$ would be precisely the face that realizes $e$ as the interface of a broken wheel contained in $S$), so let us assume for the purpose of contradiction that $a(C) > 1$. Definition \ref{def:consistent intro} implies that $C$ contains a broken wheel $B$. There are two sub-cases.

\begin{itemize}[leftmargin=*]

\medskip

\item If $e \not \subset B$, then $S$ must also contain the mirror image $B'$ of $B$. If we modify $p$ by replacing its sub-path $B$ with $B'$, then we contradict the minimality of $a(C)$.

\medskip

\item If $e \subset B$, then the interface $e'$ of the broken wheel $B$ is an edge between two vertices of the shrub $S$. If $e' \subset S$, then we contradict the minimality of $a(C)$ and the fact that $a(C) > 1$. If $e' \not\subset S$, then there is a sub-path of $p$ from the source to the tail of $e'$, and we are thus in the self-contradictory situation of item (1).

\end{itemize}

\medskip

\noindent (3) If $j \notin \{i,i'\}$, then let us choose $p,p',e$ such that $a(p \cup e, p')$ is minimal. In this case, Definition \ref{def:consistent intro} implies that one of $p \cup e$ or $p'$ contains a broken wheel $B$ whose interface is contained in $r(p \cup e, p')$. If $B \subseteq p$ or $B \subseteq p'$, then we may modify the path $p$ or $p'$ by replacing its sub-path $B$ with its mirror image, and contradict the minimality of $a(p \cup e, p')$. The only other possibility is that $e \subset B$, in which case the interface of $B$ must be an edge $e' : i' \rightarrow v$ for some vertex $v \in p$, as in Figure \ref{fig:6}.

\begin{figure}[H]
\includegraphics[scale=0.40]{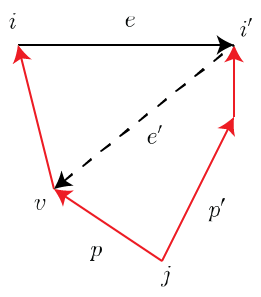} 
\caption{The situation in item (3).}
\label{fig:6}
\end{figure}

\noindent If $e' \subset S$, then concatenating $e'$ with the sub-path of $p$ that goes from $v$ to $i$ puts us in the situation of item (2) above. Meanwhile, if $e' \not\subset S$ and $v = j$, the cycle formed by $p'$ and $e'$ also puts us in the situation of item (2); since $e'$ must therefore be the interface of a broken wheel $B$ contained in $S$, replacing $p'$ by the mirror image $B'$ of $B$ would contradict the minimality of $a(p \cup e, p')$. Finally, if $e' \not\subset S$ and $v \neq j$, then we note that
$$
a(p' \cup e', p'') < a(p \cup e, p')
$$
(where $p''$ is the sub-path of $p$ that goes from $j$ to $v$) contradicts the minimality of $a(p \cup e, p')$. \end{proof}

\medskip

\begin{proof} \emph{of Proposition \ref{prop:good or bad}:} We will treat the case $s \in \{1,\dots,k-1\}$, and leave the analogous case $s = k$ as an exercise to the reader. Consider the paths $p_s$ and $p'_s$ of \eqref{eqn:path p}--\eqref{eqn:path p'}. Definition \ref{def:consistent intro} states that one of these paths must contain a broken wheel $B$; without loss of generality, let us assume that $B \subseteq p_s$. If $e_s$ were not part of $B$, then we would be able to modify $p_s$ by replacing its sub-path $B$ with its mirror image $B'$, and thus contradict the minimality of $a(p_s,p'_s)$. Therefore, we may assume that $e_s$ is part of $B$, and thus there exists $v \in p_s$ and an edge
$$
i \xrightarrow{e} v
$$
such that the region bounded by $e$ and $p_s$ is a face. If $v = j$, then the index $s$ is good (since the whole of $p_s$ is the sought-for broken wheel, and its mirror image must coincide with $p'_s$ by minimality). Otherwise $v \neq j$ and let us consider the paths
\begin{align*}
&\tilde{p}_s : j \rightarrow \dots \rightarrow v \\
&\tilde{p}'_s : j \rightarrow \dots \xrightarrow{e_{s+1}} i \xrightarrow{e} v
\end{align*}
as in Figure \ref{fig:7}. 

\begin{figure}[H]
\includegraphics[scale=0.40]{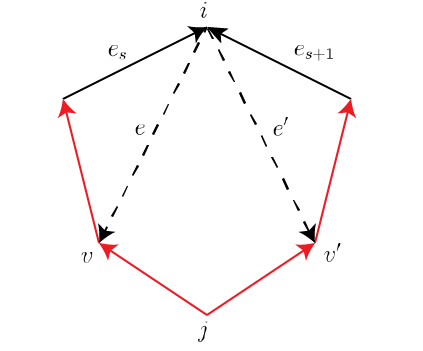} 
\caption{A bad case.}
\label{fig:7}
\end{figure}

\noindent Definition \ref{def:consistent intro} implies that one of the paths $\tilde{p}_s$ and $\tilde{p}'_s$ must contain a broken wheel $\tilde{B}$. If $\tilde{B}$ did not contain the edges $e_{s+1}$ or $e$, then we could contradict the minimality of $a(p_s,p'_s)$ by replacing $\tilde{B}$ with its mirror image $\tilde{B}'$. We are left only with the possibility of $\tilde{B}$ containing the edges $e_{s+1}$ or $e$, and we have two cases

\medskip

\begin{itemize}[leftmargin=*]

\item If the interface $e'$ of $\tilde{B}$ is an edge from $i$ to some $v' \in p'_s$,  then we assume $v' \neq j$ (as the case $v' = j$ can be treated like the case $v=j$ was treated above). We are thus in the situation of Figure \ref{fig:7} and the index $s$ is bad.

\medskip

\item If the interface $e'$ of $\tilde{B}$ is an edge from $v$ to some vertex $v' \in p'_s \backslash \{i\}$, then we are in the situation of Figure \ref{fig:8}. We have two sub-cases. If $e' \subset S$, then we contradict the minimality of $a(p_s,p'_s)$. On the other hand, if $e' \not\subset S$, then Proposition \ref{prop:no edges} forces $e'$ to be the interface of a broken wheel $\bar{B} \subset S$. The paths
$$
v' \xrightarrow{\bar{B}} v \rightarrow \dots \xrightarrow{e_s} i
$$
and $v' \rightarrow \dots \xrightarrow{e_{s+1}} i$ contradict the minimality of $a(p_s,p'_s)$.

\medskip

\begin{figure}
\includegraphics[scale=0.40]{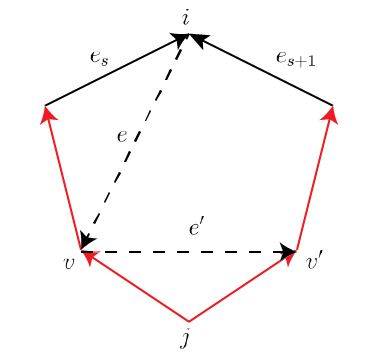} 
\caption{An impossible case.}
\label{fig:8}
\end{figure}

\end{itemize}

\end{proof}

\begin{proof} \emph{of Proposition \ref{prop:key}:} If $i$ is not an addable vertex, there must exist a bad index $s \in \{1,\dots,k\}$, i.e. either the situation of $s=1$ in the picture on the left of Figure \ref{fig:4} or the situation of $s=3$ in the picture on the right of Figure \ref{fig:4}. In both of these cases, one can see a broken wheel in $S+i$ whose mirror image is not contained in $S+i$, thus precluding $S+i$ from being a shrub.

\medskip

\noindent Conversely, suppose that $i$ is an addable vertex, and let us show that $S+i$ is a shrub. It is clear that $i$ can be reached via a path from the root, and that there are no vertices $\notin S+i$ inside the polygonal regions incident to $i$ in Figure \ref{fig:3}.

\medskip

\noindent Assume for the purpose of contradiction that $S+i$ contains the entire boundary of a face. Since $S$ cannot contain the entire boundary of a face (as $S$ is a shrub), then the boundary in question must involve the vertex $i$. However, this would require an edge from $i$ to a vertex of $S$, which is not in $S+i$ by assumption.

\medskip

\noindent Now let us assume that $S+i$ contains a broken wheel $B$, and let us show that it also contains its mirror image. Since $S$ is already a shrub, we may assume that the broken wheel $B$ involves the vertex $i$. By the definition of an addable vertex, all possible edges between $i$ and $S$ are as in Figure \ref{fig:3}. Thus, the interface of the broken wheel $B$ must be one of the dotted edges in Figure \ref{fig:3}, and it is clear that the mirror image of $B$ is also contained in $S+i$.  \end{proof}

\medskip

\begin{proof} \emph{of Proposition \ref{prop:count}:} If $i$ is addable to $S$, then all $s \in \{1,\dots,k\}$ are good. Therefore, there exist only $k-1$ outgoing edges from $i$ to $S$, and they are arrayed as in Figure 3. Among any family of faces passing through $i$ and without other pairwise intersections, no two faces can pass through the same outgoing edge, so the cardinality of the family is at most $k-1$. It is also easy to see that this maximum can be achieved, by taking for instance the collection of faces incident to $e_1,\dots,e_{k-1}$ in clockwise order around $i$. 

\medskip

\noindent If $i$ is non-addable to $S$, then there exists a bad index $s$. Assume first that $s \in \{1,\dots,k-1\}$, e.g. we are in the situation of $s=1$ in the picture on the left of Figure \ref{fig:4}. The two faces contained in the region $r_s$, together with the faces incident to $e_1,\dots,e_{s-1}$ in clockwise order around $i$, and the faces incident to $e_{s+2},\dots,e_k$ in counterclockwise order around $i$, yield altogether a family of $k$ faces which only pairwise intersect at $i$.

\medskip

\noindent If $s = k$ is a bad index, then we are in the situation in the picture on the right of Figure \ref{fig:4}. Without loss of generality, let us assume that there is a face incident to $e_k$ in clockwise order around $i$. Then this face together with the faces incident to $e_1,\dots,e_{k-1}$ in clockwise order around $i$, yield the required family of $k$ faces which only pairwise intersect at $i$. \end{proof}

\end{document}